\documentclass[reprint,amsmath,amssymb,apr,aip,onecolumn,nofootinbib]{revtex4-2}
\usepackage{graphicx}
\usepackage{graphics}
\usepackage[T1]{fontenc}
\usepackage{mathptmx}
\usepackage{times}
\usepackage{amsmath}
\usepackage{amssymb}
\usepackage{dcolumn}
\usepackage{color}
\usepackage{bm}
\usepackage{units}
\usepackage{booktabs}
\usepackage{footnote}
\usepackage{multirow}
\usepackage{tabularray}
\UseTblrLibrary{booktabs}
\usepackage{threeparttable}
\usepackage[strict]{changepage}
\usepackage{algorithm}
\usepackage{algpseudocode}
\usepackage{framed}

\definecolor{formalshade}{rgb}{0.95,0.95,1}

\usepackage{acronym}
\acrodef{AI}[AI]{Artificial Intelligence}
\acrodef{DL}[DL]{Deep Learning}
\acrodef{PIM}[PIM]{Processing-In-Memory}
\acrodef{IoT}[IoT]{Internet-of-Things}
\acrodef{HWA}[HWA]{Hardware-Aware}
\acrodef{ML}[ML]{Machine Learning}
\acrodef{IMC}[IMC]{In-Memory Computing}
\acrodef{PCM}[PCM]{Phase Change Memory}
\acrodef{PGD}[PGD]{Projected Gradient Descent}
\acrodef{DNN}[DNN]{Deep Neural Network}
\acrodef{RRAM}[RRAM]{Resistive Random-Access Memory}
\acrodef{ADC}[ADC]{Analog-to-Digital Converter}
\acrodef{DAC}[DAC]{Digital-to-Analog Converter}
\acrodef{MVM}[MVM]{Matrix-Vector Multiplication}
\acrodef{AIMC}[AIMC]{Analog In-Memory Computing}
\acrodef{DSE}[DSE]{Design Space Exploration}
\acrodef{CNN}[CNN]{Convolutional Neural Network}
\acrodef{FGSM}[FGSM]{Fast Gradient Sign Method}
\acrodef{NLP}[NLP]{Natural Language Processing}
\acrodef{MAC}[MAC]{Multiply and Accumulate}
\acrodef{MVM}[MVM]{Matrix-Vector Multiplication}
\acrodef{NVM}[NVM]{Non-Volatile Memory}
\acrodef{GBDA}[GBDA]{Gradient-based Distributional Attack}
\acrodef{ASR}[ASR]{Adversarial Success Rate}
\acrodef{RGB}[RGB]{Red Green Blue}
\acrodef{GPU}[GPU]{Graphics Processor Unit}
\acrodef{CMOS}[CMOS]{Complementary Metal-Oxide Semiconductor}
\acrodef{ART}[ART]{Adversarial Robustness Toolbox}
\acrodef{PTQ}[PTQ]{Post Training Quantization}
\acrodef{LDPU}[LDPU]{Local Digital Processing Unit}
\acrodef{GDP}[GDP]{Gradient Descent Programming}
\acrodef{RTN}[RTN]{Random Telegraph Noise}
\acrodef{FP32}[FP32]{Floating-Point 32-bit}

\begin{document}
\title{The Inherent Adversarial Robustness of Analog In-Memory Computing}

\author{Corey Lammie}\email{corey.lammie@ibm.com}\affiliation{IBM Research Europe, 8803 R\"{u}schlikon, Switzerland}
\author{Julian B\"{u}chel}\affiliation{IBM Research Europe, 8803 R\"{u}schlikon, Switzerland}
\author{Athanasios Vasilopoulos}\affiliation{IBM Research Europe, 8803 R\"{u}schlikon, Switzerland}
\author{Manuel Le Gallo}\affiliation{IBM Research Europe, 8803 R\"{u}schlikon, Switzerland}
\author{Abu Sebastian}\email{ase@zurich.ibm.com}\affiliation{IBM Research Europe, 8803 R\"{u}schlikon, Switzerland}

\date{\today}
\begin{abstract}
A key challenge for \ac{DNN} algorithms is their vulnerability to adversarial attacks. Inherently non-deterministic compute substrates, such as those based on \ac{AIMC}, have been speculated to provide significant adversarial robustness when performing \ac{DNN} inference. 
In this paper, we experimentally validate this conjecture for the first time on an \ac{AIMC} chip based on \ac{PCM} devices. 
We demonstrate higher adversarial robustness against different types of adversarial attacks when implementing an image classification network. 
Additional robustness is also observed when performing hardware-in-the-loop attacks, for which the attacker is assumed to have full access to the hardware. A careful study of the various noise sources indicate that a combination of stochastic noise sources (both recurrent and non-recurrent) are responsible for the adversarial robustness and that their \emph{type} and \emph{magnitude} disproportionately effects this property.
Finally, it is demonstrated, via simulations, that when a much larger transformer network is used to implement a \ac{NLP} task, additional robustness is still observed.
\end{abstract}
\maketitle

\acresetall

\acp{DNN} have revolutionized \ac{ML} and forms the foundation for much of modern \ac{AI} systems. However, they are susceptible to a number of different types of adversarial attacks, which aim to deceive them by giving them false information. These include poisoning-, extraction-, and evasion-based attacks~\cite{Y2018akhtarIEEEAccess}. Evasion-based attacks target \ac{DNN} models by generating \emph{adversarial} inputs, which achieve a desired (usually malicious) outcome when inferred~\cite{PITROPAKIS2019100199}. For example, carefully crafted adversarial inputs can consistently trigger an erroneous classification output from a \ac{DNN} model. Hence, there is significant interest in developing algorithms and underlying compute substrates that make \acp{DNN} more robust to adversarial attacks~\cite{GhaffariLaleh2022,Woods2019,Pereira2019}.

One such compute substrate which shows significant promise for adversarial robustness is that based on \ac{AIMC}~\cite{Y2018ielminiNatElec,Sebastian2020,Y2022lanzaScience}. These architectures were primarily motivated by the need to \emph{minimize} the amount of data transferred between memory and processing units during the execution of \ac{MAC}-dominated workloads, to avoid the well known memory-wall problem~\cite{Y2018zidanNatElec,Mehonic2020}.
\ac{DNN} inference workloads are mostly comprised of \ac{MAC} operations. Consequently, when \ac{AIMC}-based hardware is used for their acceleration, their energy-efficiency and latency can be greatly improved.
For example, in an \ac{AIMC}-based accelerator based on nanoscale \ac{NVM} devices, synaptic unit-cells comprising one or more devices can be used to encode weights of \ac{DNN} layers, and the associated \acp{MVM} operations can be performed in-place by exploiting Kirchhoff's circuit laws.
When multiple crossbar arrays with physically stationary synaptic weights and digital computing units are connected using a fabric able to route data, it is possible to realize complete (end-to-end) inference workloads.
A key drawback of these accelerators is that they typically exhibit a notable degradation in accuracy compared to their full precision counterparts, on account of device-level noise and circuit non-idealities. However, \ac{HWA} training, where the \ac{DNN} is made robust via the injection of weight noise during the training process, has been found to recover much of the accuracy loss~\cite{Joshi2020,Y2023raschNatComm}. 

One often overlooked benefit of the stochasticity associated with \ac{AIMC} is robustness against adversarial attacks. In fact, prior work has demonstrated that even the robustification of \ac{DNN} models introduced by \ac{HWA} training could provide a certain level of adversarial robustness~\cite{bhattacharjeeRethinkingNonidealitiesMemristive2021,Y2021royDAC,taoNoiseStabilityRobustness2022,cherupallyImprovingAccuracyRobustness2022}. 
It has also been demonstrated that simulated hardware non-idealities yield adversarial robustness to mapped \acp{DNN} without any additional optimization~\cite{bhattacharjeeRethinkingNonidealitiesMemristive2021} and simulated adversarial attacks crafted without the knowledge of the hardware implementation are less effective in both black- and white-box attack scenarios~\cite{Y2021royDAC,taoNoiseStabilityRobustness2022}.
Novel \ac{DNN} training schemes have been presented~\cite{cherupallyImprovingAccuracyRobustness2022}, which improve robustness when evaluated using a number of simulated \acp{DNN} with measured noise data obtained from three different \ac{RRAM}-based \ac{IMC} prototype chips.
Additionally, simulations have been used to investigate the relationships between adversarial robustness, \ac{HWA} training, non-idealities in crossbars, analog implementation of activation functions, and supporting circuits~\cite{paudelImpactOnchipTraining2022}.
Finally, attacks have been proposed to generate adversarial examples leveraging the non-ideal behavior of \ac{RRAM}-based devices~\cite{shangFaultAwareAdversaryAttack2022,mclemoreExploitingDevicelevelNonidealities2023,lvVADERLeveragingNatural2021,lvVariationEnhancedAttacks2023a}.
However, none of these studies have been verified on real \ac{AIMC}-based hardware. Moreover, it is not well understood, to what degree, the different noise sources contribute to adversarial robustness. 

In this paper, we investigate the inherent adversarial robustness, i.e., how inherently vulnerable a system is to various types of attack vectors, of \ac{AIMC} for (i) a Resnet-based \ac{CNN} trained for image classification (CIFAR-10), and (ii) the pre-trained RoBERTa\cite{DBLP:journals/corr/abs-1907-11692} transformer network fine-tuned for a pair-based sentence textual entailment \ac{NLP} task (MNLI GLUE)\cite{williams-etal-2018-broad}.

\color{black}
Adversarial \ac{ML} is a broad research field~\cite{Chakraborty2021}. While we confine the scope of this paper to the investigation of the inherent adversarial robustness of \ac{AIMC}-based hardware against evasion-based adversarial attacks, there are many other different types of attack strategies which exist.
Instead of generating \emph{adversarial} inputs, poising- and back-door attacks corrupt models by modifying the data used to train them~\cite{Biggio2012,Wenger_2021_CVPR}.
Side-channel~\cite{Standaert2010} and other other types of attacks which take advantage of physical characteristics of systems~\cite{Wang2024}, can be used to infer hidden information about a model and introduce malicious instability.
\color{black}

Simulation (for both networks) and real hardware experiments (for the Resnet-based \ac{CNN}) are performed (see Methods) using an \ac{AIMC} chip based on \ac{PCM} devices.
First, for the Resnet-based \ac{CNN}, in agreement with prior findings, we observe that injecting noise during \ac{HWA} training improves robustness to adversarial attacks. 
We further observe, that when \ac{HWA} trained networks are deployed on-chip, this robustness is notably improved. 
It is investigated what characteristics of different noise sources contribute to this robustness, and demonstrated that the \emph{type} and \emph{magnitude} properties of stochastic noise sources is much more critical than their recurrence property.
We consider the type as a binary attribute indicating whether, for a given noise source, its magnitude is determined as a function of the input.
Additionally, we investigate the efficacy of hardware-in-the-loop attacks, and demonstrate that even these kind of attacks are less effective when stochastic systems are targeted.
Finally, using the RoBERTa transformer network, it is demonstrated through simulation, that for this much larger network and different input modality, additional adversarial robustness is still observed.

\section*{Results}
\subsection*{Experimental validation of adversarial robustness}\label{sec:hwexp}
To experimentally study the adversarial robustness, we employed a \ac{PCM}-based \ac{AIMC} chip with tiles comprising $256\times 256$ synaptic unit-cells \cite{Y2023legalloNatElec}.
Each unit-cell \textcolor{black}{contains} four \ac{PCM} devices (see Fig. \ref{fig:concept}\textbf{b}).
The weights obtained via \ac{HWA} training are stored in terms of the analog conductance values of the \ac{PCM} devices, where two devices are used to store the positive and negative weight components, respectively.
The conductance variations associated with these conductance values are thought to be the primary reason for potential adversarial robustness.
As shown in Fig. \ref{fig:concept}\textbf{a}, the intrinsic stochasticity associated with the conductance variations is likely to make the design of an adversarial attack rather difficult. 
The conductance variations themselves fall into two different categories. There is a non-recurrent category that results from the inaccuracies associated with programming a certain analog conductance value. This is typically referred to as programming noise \cite{Y2023legalloNatElec}. Besides this non-recurrent noise component, there is a recurrent variation in the conductance values arising from $1/f$ noise \cite{Y2009nardonePRB} and \ac{RTN} characteristic of \ac{PCM} devices. In subsequent sections, we will investigate the role of the recurrent and non-recurrent noise sources with respect to adversarial robustness.

For this study\textcolor{black}{,} we consider three different types of adversarial attacks and five different target platforms.
The different types of attacks are (i) \ac{PGD}~\cite{madryDeepLearningModels2019}, (ii) Square~\cite{andriushchenkoSquareAttackQueryefficient2020}, and (iii) OnePixel~\cite{suOnePixelAttack2019}.
The attacks range from targeting small (localized) to larger (for some attacks, entire) input regions.
Specifically, the \ac{PGD} attack was chosen, as it is equivalent to the \ac{FGSM} attack when the starting point is not randomly chosen and the $L_{\infty}$ norm is used~\cite{huangBridgingPerformanceGap2022}. Hence, it is superior to \ac{FGSM} when more than one iteration is used to generate adversarial inputs.
The Square and OnePixel attacks do not rely on local gradient information, and thus, are not affected by gradient masking, which is a defensive strategy aimed at diminishing the efficacy of gradient-based attacks by obfuscating the model's loss function, rendering it less informative or harder to optimize\cite{tramèr2020ensemble}.
The OnePixel attack only permutes a small region (i.e., a single pixel) of the input, rather than a larger region, as typically targeted by other attacks.

The five different attack target platforms considered are: (i) the original floating-point model (Original Network (FP32)), (ii) the floating-point \ac{HWA} retrained model (HWA Retrained (FP32)), (iii) a digital hardware accelerator with 4-bit fixed-point weight precision and 8-bit activation precision (Digital), (iv) a \ac{PCM}-based \ac{AIMC} chip model\footnote{The \ac{PCM}-based \ac{AIMC} chip model is described in the Methods section and a match between the model and experimental data is provided in Supplementary Note 1.} (AIMC Chip Model), and (iv) a \ac{PCM}-based \ac{AIMC} chip (AIMC Chip).
For the first target, an existing pre-trained model is used.
For the second target, the pre-trained model is re-trained using \ac{HWA} training. For the third, fourth, and fifth targets, the parameters of the re-trained model are mapped to the corresponding hardware and deployed\footnote{All auxiliary operations are performed in floating-point precision.}.

To determine the effectiveness of adversarial attacks, different evaluation metrics can be used.
Typically, the clean (baseline) accuracy is directly compared to the accuracy when adversarial inputs are inferred.
In this paper, we adopt the \ac{ASR} metric \cite{DBLP:journals/corr/abs-1902-06705}, which considers only samples that are classified correctly by the network.
More specifically, the \ac{ASR} is determined as follows. First, model predictions are determined for both clean and adversarial inputs. Using target labels of the clean inputs, the incorrectly classified clean inputs are identified. These target labels and the predictions of adversarial inputs (adversarial predictions), which originated from the incorrectly classified clean inputs,  are then masked and discarded.
Traditional \ac{ML} evaluation metrics can then be computed using the masked target labels and adversarial predictions.
We concretely describe the determination of the \ac{ASR} in terms of accuracy using Algorithm \ref{alg:cap} (see Methods).

By adopting the \ac{ASR} metric, we can evaluate and compare the performance of adversarial attacks which are \emph{generated} and/or \emph{evaluated} using different target platforms. To demonstrate this ability, in Extended Data Table \ref{independence_table}, we compute both the test set accuracy and \ac{ASR} during training for a floating-point Resnet-based \ac{CNN} trained for CIFAR-10 image classification\footnote{\textcolor{black}{In Supplementary Note 3, we perform simulations for a more challenging image classification task.}} using two different types of attacks\footnote{These attacks are considered in all fore-coming experiments, and are described in greater detail below.}, \ac{PGD} and OnePixel, with different parameters. For the \ac{ASR} to be truely agnostic to the target platform, it should be independent of the network accuracy. 
It is observed that the \ac{ASR} is \textit{sufficiently} independent of the test set accuracy. It is only notably perturbed for extreme scenarios (i.e., when the test set accuracy is <25\%).

For each type of attack, we vary two parameters. The first parameter relates to the number of attack iterations, whereas the second parameter relates to the attack magnitude, i.e., the maximum amount of distortion the attack can introduce to the input. 
When evaluating the \ac{ASR} as a function of different attack parameters, especially for those relating to magnitude, caution must be taken so that the ground truth label is not changed by extreme permutations.
If the severity of an attack or the number of attack iterations is too high, then the ground truth labels of the generated adversarial inputs can differ from the original inputs, which is problematic as the underlying semantics of the original input have now changed.
Consequently, for all experiments, the maximum value for each parameter is determined by manually inspecting adversarial examples and determining the points at which the ground truth label is changed (see Methods).

To comprehensively compare the adversarial robustness for all target platforms, while considering the number of attack iterations and magnitude, a contour or 3D plot, per target platform, is required.
These are difficult to compare relative to each other.
Hence, instead, we generate an ASR for each target platform, which is indicative of the \emph{average} behavior of the objective as a function of both parameters.
This is generated by projecting a straight line from the bottom-left to top-right corners of each contour plot. Along this line, the \ac{ASR} is extracted.
Each point in this line is associated with two values: one governing the number of attack iterations and another governing the attack magnitude. A dual x-axis is used to associate these.

For each evaluation platform, aside from the \ac{AIMC} chip, the strongest attack scenario is considered, i.e., attacks are both generated and evaluated using the same platform. To both generate and evaluate attacks on the \ac{AIMC} chip, hardware-in-the-loop attacks would have to be performed -- such attacks are non-trivial to perform on stochastic hardware, and hence, instead, are the sole focus of \textcolor{black}{the Hardware-in-the-loop adversarial attacks section}. Consequently, for the \ac{AIMC} chip, adversarial inputs are generated using the \ac{AIMC} chip model.
For the Software (FP32) evaluation platform, attacks are generated using the same platform with FP32 and HWA Retrained (FP32) weights, respectively.

In Fig.~\ref{fig:hwexp}\textbf{a-c}, the \ac{ASR} quantity is related to both aforementioned parameters for the \ac{AIMC} chip evaluation platform.
As can be observed in Fig.~\ref{fig:hwexp}\textbf{d-f}, the envelope for the floating-point \ac{HWA} retrained model is smaller than the original floating-point model, meaning it is more robust.
This result is consistent with prior work~\cite{Y2021royDAC,bhattacharjeeRethinkingNonidealitiesMemristive2021,cherupallyImprovingAccuracyRobustness2022,droletHardwareawareTrainingTechniques2023}.
The digital hardware accelerator has an even smaller envelope, meaning it is more robust than the floating-point \ac{HWA} retrained model. It is noted that like the operation of the digital hardware accelerator, the evaluation of the floating-point \ac{HWA} retrained model is deterministic. \textcolor{black}{Contrastingly,} the digital hardware platform is subject to quantization noise and accumulation error~\cite{9794618}.
While this error is introduced during normal system operation, as digital hardware is typically deterministic (as is assumed here), it can be predicted in advance, i.e., when adversarial inputs are generated\footnote{In fact, it has been demonstrated that adversarial attacks targeting these deterministic systems can be more effective~\cite{DBLP:journals/corr/abs-1810-00208}.}. Critically, (i) there is a good agreement between the modelled ASR for the AIMC chip and the ASR rate of the hardware models on the \ac{AIMC} chip, and (ii) the hardware experiments on the \ac{AIMC} chip result in the smallest envelopes, and hence, the highest level of robustness to all investigated adversarial attacks.

\subsection*{Source of adversarial robustness}\label{sec:source}
In this section, we present a detailed study of the various noise sources that contribute to the higher adversarial robustness observed experimentally on the \ac{AIMC} chip (see Methods).
Specifically, we investigate the role of four different noise properties -- the recurrence (or lack thereof), location, type, and magnitude.
For this study, we rely on the hardware model of the \ac{AIMC} chip.
As described earlier, non-recurrent noise sources introduce noise \emph{once} during operation, i.e., when weights are programmed, whereas recurrent noise sources introduce noise \emph{multiple times} during operation, i.e., when a \ac{MVM} operation is performed.
In the \ac{AIMC} chip model, the recurrent noise sources are sampled multiple times (specifically per mini-batch), whereas non-recurrent noise sources are only sampled once.
All noise sources are assumed to be Gaussian and centered around zero.

The non-recurrent component of stochastic weight noise primarily arises from programming noise. The recurrent components of the stochastic weight noise primarily arises from both device read noise ($1/f$ and RTN) and output noise, which is introduced at the crossbar output, before currents are read out using \acp{ADC}.
Output noise includes $1/f$ noise from amplifiers and other peripheral circuits, as well as other circuit non-linearities (IR drop, sneak paths, etc.) of small magnitude that can be approximated as a random perturbation on the crossbar output.
Hence, in the model, non-recurrent weight noise is modelled, and at the output, these combined effects are modelled using additive recurrent noise of a fixed magnitude which is assumed to be independent of the total column conductance.
For each configuration, other deterministic noise sources, e.g., input and output quantization, are also modelled.
Other non-deterministic noise sources are assumed to contribute negligibly\footnote{\textcolor{black}{As evidenced by the strong match between the AIMC chip and the AIMC chip model across all hardware
experiments.}}, and hence, are not modelled.

In order to assess how three properties of stochastic noise sources\footnote{\textcolor{black}{In Supplementary Note 4, we investigate how the inherent adversarial robustness of AIMC-based hardware is effected by temporal drift.}}, namely, recurrence, location, and type affect adversarial robustness, we modify the behavior of the \ac{AIMC} chip model to focus exclusively on a single stochastic noise source.
Separately, we investigate both recurrent and non-recurrent noise sources at two locations: the weights, whose effect is determined as a function of the input, and at the output, whose effect is input-independent. A total of four stochastic noise sources are considered: (i) recurrent output noise, (ii) recurrent weight noise, (iii) non-recurrent weight noise, and (iv) non-recurrent output noise.
The effect of the noise magnitude is determined by, for each noise source, modifying the noise magnitude such that the resulting test set accuracy is lower than the floating-point test set accuracy by a desired percentage (i.e., drop).
We consider drops of 5\% and 10\%.

In Fig.~\ref{fig:source}\textbf{a}, for a large number, $n$=1,000, of repetitions, the test set accuracy is reported for both non-recurrent and recurrent output and weight noise\footnote{\textcolor{black}{In Supplementary Notes 2 and 5, we investigate adversarial robustness to varying degrees of stochasticity and combinations of output and weight noise.}}.
It is observed that, for non-recurrent and recurrent noise sources normalized to produce the same test set accuracy, the noise magnitude is approximately equal, i.e., the \emph{recurrence} property does not effect the resulting \emph{average} test set accuracy. For non-recurrent noise sources the variation of the resulting test set accuracy is larger.
For noise sources normalized to produce a larger test set accuracy drop (10\%), the noise magnitude is larger, compared to the smaller drop (5\%).
Intuitively, as the noise magnitude is increased, the test set accuracy decreases and the network becomes more robust to adversarial attacks.

To investigate the effect of the recurrence, location, and type properties on adversarial robustness, further experiments are performed in Fig.~\ref{fig:source}\textbf{b}. Comparisons are made using one of the selected attacks, \ac{PGD}.
For the noise magnitudes associated with the smaller test set accuracy drop (5\%), the \ac{ASR} envelope is determined for $n$=10 repetitions, for the four aforementioned noise sources.
Two key observations can be made: (i) output noise exhibits greater adversarial robustness compared to weight noise, and (ii) in addition to not effecting the average test set accuracy, the recurrence property does not effect the \ac{ASR}.
We postulate that weight noise leads to less robustness when normalized to produce the same error, as the effect of weight noise is input dependent, whereas output noise is not.

In Fig.~\ref{fig:source}\textbf{c-e}, we further compare the \ac{ASR} envelope for the \ac{PGD}, Square, and OnePixel attacks. It is observed that, on average, the model with only output noise exhibits the highest adversarial robustness. The model with only weight noise is the least robust. The \ac{AIMC} chip model and \ac{AIMC} chip, with test set accuracy values of 84.85\% and 84.31\%, respectively, i.e., drops of 3.57\% and 4.11\%, exhibit greater adversarial robustness than the modified \ac{AIMC} chip model with only weight noise, but less adversarial robustness than the modified \ac{AIMC} chip model with only output noise.
From this analysis, it can be concluded that, the type and magnitude properties of stochastic noise sources have the greatest influence on adversarial robustness. The location and recurrence properties have negligible influence.

\subsection*{Hardware-in-the-loop adversarial attacks}\label{sec:hwloop}
Next, we determine the efficacy of hardware-in-the-loop attacks, i.e., where it is assumed that the attacker has full access to the \ac{AIMC} chip.
We compare the efficacy of one white- and one black-box attack (PGD and OnePixel).
White box attacks are especially difficult to perform for stochastic hardware, as the construction of representative hardware models (with minimal mismatch) is non-trivial, and in many cases, even unfeasible.
While automated \ac{ML}-based in-the-loop modelling~\cite{Y2020chakrabortyDAC} approaches can be utilized, they require a significant amount of data, which is instance specific. Hence, they are not considered in this paper.

For hardware-in-the-loop attacks, when a white-box attack is deployed, to perform backwards propagation, for each layer, weights and cached inputs are required~\cite{akhtarThreatAdversarialAttacks2018}.
Additionally, the output(s) of the network is(are) required.
As ideal, i.e., floating-point precision, weights cannot be programmed, there is some deviation between the target and programmed conductances, so the target weights (if known) cannot simply be used by the attacker. Additionally, read noise introduces random fluctuations when \ac{MVM} workloads are executed. Representative weights can be inferred by solving the following optimization problem, as described in~\cite{jub}
\begin{equation}
    \mathbf{\hat{G}} = \operatorname*{argmin}_{\mathbf{\hat{G}}} \sum_{b=1}^{B} \|\mathbf{\hat{G}}\mathbf{x}_b - \tilde{\mathbf{y}}_b\|_2.
\end{equation}

Inputs to each layer can be cached during normal operation by probing input traces to \acp{DAC}. To perform backwards propagation, this information can be used to construct a proxy network graph (see Methods), for which gradients can be computed in floating-point precision using the chain-rule.
For sake of practicality, an ideal backwards pass is assumed, i.e., straight-through-estimators are used for regions which are non-differentiable, and all values are assumed to be in floating-point precision.
It is noted that, as the next candidate adversarial input to the network is usually dependent on the the result of backwards propagation of the previous candidate input, this process cannot normally be pipelined.
Consequently, depending on the operation speed of the chip, attacks generated using \ac{AIMC} chips can be susceptible to low-frequency noise and temporal variations, such as conductance drift~\cite{Y2022yangSmallScience}.
To mitigate these effects, as reprogramming all devices after each adversarial example is presented is not desirable, $\mathbf{\hat{G}}$ can be re-inferred.
It is noted that, for black box attacks, these effects cannot be effectively mitigated - except in the scenario where the attacker is aware of exactly when the chip was programmed.

In Fig.~\ref{fig:hwloop}, we compare the efficacy of hardware-in-the-loop attacks generated and evaluated using the AIMC chip to three different attack scenarios, where: (i) adversarial inputs are generated and evaluated using the digital hardware accelerator, (ii) adversarial inputs are generated and evaluated using the AIMC chip model, and (iii) adversarial inputs are generated using the AIMC chip model and evaluated using the AIMC chip.
When evaluated using the AIMC chip, adversarial attacks generated using hardware-in-the-loop attacks on the AIMC chip are more effective than on the AIMC chip model.
Hardware-in-the-loop attacks generated and evaluated on the AIMC chip model are marginally less effective.
Critically, for all scenarios involving either the AIMC chip or AIMC chip model, additional adversarial robustness is observed compared to hardware-in-the-loop attacks both generated and evaluated on the digital hardware platform.

We emphasize, that to take the required circuit measurements to perform hardware-in-the-loop attacks, even for black-box attacks, where the output logits are needed, significant knowledge about the underlying hardware is required.
In some instances, direct traces may be unavailable, meaning that a combination of other signals must be used as a proxy, reducing attack efficacy.
In others, developing a realistic hardware model is too time consuming and not practically feasible.
Similarly to when a representative hardware model is attacked, unlike for deterministic systems, e.g., digital hardware accelerators, the generated adversarial inputs are specific to that particular instance of the hardware, and hence not useful for large-scale attacks targeting many instances at the same time.

\subsection*{Applicability to transformer-based models and natural language processing tasks}
Finally, to determine whether additional adversarial robustness is still observed for much larger transformer models and different input modalities, we simulate a pre-trained floating-point RoBERTa model fine-tuned for one GLUE task, MNLI. The RoBERTa model has approximately 125M parameters and the MNLI task comprises 393K training and 20K test samples (pairs of sentences).
When fine-tuning the model for the down-stream MNLI task, \ac{HWA} training was performed (see Methods).
This model exceeds the weight capacity of the IBM HERMES Project Chip, so instead of performing hardware experiments, simulations were conducted using the \ac{AIMC} chip model.

A number of different hardware attacks for \ac{NLP} tasks with text-based input exist.
These usually target either specific characters, words, or tokens~\cite{Y2020zhangACMTIST}.
One major challenge when generating adversarial attacks for \ac{NLP} tasks is semantic equivalence.
For images, small permutations are usually not perceivable. However, for text, small permutations are much more notable (even at a character level), and are more likely to alter the semantics of the input.
We consider the \ac{GBDA} attack~\cite{DBLP:journals/corr/abs-2104-13733}, which instead of constructing a single adversarial example, as done by most types of attacks, searches for an adversarial \emph{distribution} and considers semantic similarity using the BERTScore metric~\cite{DBLP:journals/corr/abs-1904-09675}.
This enforces some degree of semantic equivalence. In Extended Data Table~\ref{table:prompts}, we list a number of adversarial text-based inputs for different $\lambda_{sim}$ and $n_{iters}$ parameter values, in addition to their respective BERTScore values. 

Similarly to as done in Section~\ref{sec:hwexp}, we consider varying two attack parameters, which relate to the number of attack iterations and the magnitude, $n_{iter}$ and $\lambda_{sim}$, respectively\footnote{The $\lambda_{sim}$ parameter is inversely proportional to the attack magnitude.}. 
In Fig.~\ref{fig:transformer}, the \ac{ASR} is reported for the first four target platforms (listed in Section~\ref{sec:hwexp}).
Additional robustness is once again demonstrated for the \ac{AIMC} chip model for a \emph{range} of attack parameter values and BERTScore values. 
This is significant, as it indicates that additional adversarial robustness is still observed for (i) different input modalities, and (ii) much larger and more parameterized networks.

\section*{Discussion}
Adversarial attacks and other types of cyber attacks pose a significant threat to \ac{DNN} models.
Hence, a number of dedicated defence mechanisms~\cite{Y2023wangIEEEComm} and training methodologies~\cite{baiRecentAdvancesAdversarial2021} have been proposed.
The most effective defense mechanism, which is adversarial training, is too computationally expensive for practical deployment\cite{REN2020346}.
Moreover, heuristic-based defenses have been demonstrated to be vulnerable to adaptive white-box adversaries. Therefore, their widespread adoption is limited and their use is not often considered.
In both black- and white-box settings, when the attacker does not have unrestricted access to the hardware, we have demonstrated, empirically, both non-recurrent and recurrent stochastic noise sources influence robustness to adversarial attacks.
These stochastic noise sources, which are present in \ac{AIMC} chips, inherently act as an effective defence mechanism against adversarial attacks.
When the attacker has unrestricted access to the hardware, we demonstrate that in addition to being much more difficult to effectively attack, inherent adversarial robustness is also observed.

For \ac{AIMC} chips, this increased robustness comes at no additional hardware cost, and does not require any augmentation to the training and deployment pipelines typically employed for \acp{DNN} to be accelerated using this hardware.
Moreover, the inherent stochasticity of these chips makes them difficult to target using hardware-in-the-loop based attacks.
While other dedicated defence mechanisms may be more effective, they incur some additional hardware/resource cost, and most critically, require modification to training and/or deployment pipelines. This adversarial robustness highlights yet another powerful computational benefit of the intrinsic stochasticity associated with AIMC chips similar to 
prior demonstrations such as in-memory factorization \cite{Y2023langeneggerNatNano}, bayesian \cite{Y2023harabiNatElec}, and combinatorial optimization \cite{Y2020caiNatElec}.

Looking forward, as device and circuit technologies, and strategies for the mitigation of stochastic behavior continue to improve, it is expected that the inherent adversarial robustness of these chips will decrease.
We demonstrate, that, particular types of stochastic noise sources, which introduce a relatively small additional error, at no additional resource cost, can be used to effectively defend against adversarial attacks.
To improve adversarial robustness, for future \ac{AIMC} designs, the presence, magnitude, and location of these noise sources, could be considered.
For hardware accelerators based on other technologies, these types of noise sources could intentionally be introduced, at the cost of a small degradation in accuracy, in lieu, or in conjunction with, other adversarial defence mechanisms.

\color{black}
To conclude, in this paper, we performed hardware experiments using an \ac{AIMC} chip based on \ac{PCM} devices. Additionally, we performed simulations grounded by experimental data gathered from this chip to determine what characteristics of different noise sources contributed to this robustness, and evaluated the adversarial robustness of larger networks with different input modalities.
To perform all these experiments, we developed a standardized and extendable methodology to evaluate the adversarial robustness of \ac{AIMC} chips.
With little effort, the \ac{ASR} metric and our attack generation and evaluation methodology can be repurposed to evaluate the adversarial robustness against other attack types, other types of \ac{AIMC} hardware, e.g., \ac{RRAM}-based \ac{AIMC} chips, and chips with a different underlying architecture.
All types of \ac{NVM} devices are susceptible to some degree of conductance drift, meaning that as long as \ac{NVM} devices are used, interleaving samples during evaluation will counteract the effects of drift. The optimization problem formulated in Eq. 1 can also be repurposed for any arbitrary programmable conductive device, meaning that our hardware-in-the-loop attack methodology can also be applied to other types of \ac{AIMC} hardware, or with more effort, to other emerging technologies.

In addition to evaluating the adversarial robustness of different \ac{AIMC} hardware configurations and technologies, future work entails the investigation different attack types, e.g., poisoning-based attacks, and physical-based attacks, e.g., those which maliciously introduce instability to the system. The interaction of inherent adversarial robustness and stochastic dropouts~\cite{Krestinskaya2019,Krestinskaya2020} is also another interesting future research direction.
\color{black}

\newpage

\newpage
\color{black}
\section*{References}
\bibliography{References}

\newpage
\section*{Methods}
\subsection*{Datasets and Neural Network Models} \label{methods:datasets}
Two datasets were used for evaluation: (i) CIFAR-10, an image classification task comprising 50,000 training images and 10,000 test images (32 $\times$ 32 pixel \ac{RGB}), and (ii) the NMLI corpus, which is comprised of 443K sentence pairs annotated with textual entailment information. These are labelled as either \emph{neutral}, \emph{contradiction}, or \emph{entailment}. For image classification, standard input pre-processing steps were performed to normalize inputs to zero-mean with a single standard deviation. The training set was used to construct a validation and training set for training the ResNet9 network, and the test set was used for evaluation.
For sentence pair annotation, standard pre-processing and tokenization steps were performed. The pre-trained RoBERTa transformer network, which was  originally trained in floating-point precision using
five English-language corpora of varying sizes and domains, totaling over 160GB of uncompressed text~\cite{DBLP:journals/corr/abs-1907-11692}, was fine-tuned using the matched MNLI validation set and evaluated using the mismatched MNLI validation set.

\subsection*{Neural Network Model Training}
First, both neural network models were trained in floating-point precision. The ResNet9S network was trained from randomly-initialized weights using the CIFAR-10 training set. This was split into training and validation subsets with 40,000 and 10,000 samples each, respectively. A batch size of 128 was used with a cosine-annealing learning rate schedule. An initial learning rate of 0.1 and momentum value of 0.7 were used.
The pretrained-tuned RoBERTa model was fined-tuned using the matched MNLI validation set with a maximum sequence length of 256, a fixed learning rate of 2E-5, and a batch size of 16 for 10 epochs.

Next, for both the low-precision fixed-point digital and \ac{PCM}-based \ac{IMC} accelerators, hardware-aware retraining was performed.
The IBM AIHWKIT was used to inject noise during forward propagations. We refer the reader to~\cite{10.1063/5.0168089} for a comprehensive tutorial on HWA training using IBM AIHWKIT. The \texttt{InferenceRPUConfig()} preset was used with the
PCMLikeNoiseModel phenomenological inference model. The following additional modifications were made to the simulation configuration: (i) Biases were assumed to be digital (i.e., they are not encoded on the last column of \ac{AIMC} tiles).
(ii) Channel- (i.e., column-) wise weight scaling, which has a negligible performance impact, was performed. Weights were mapped to a per-channel conductance range during learning after each mini-batch.
(iii) The size of each \ac{AIMC} tile was assumed to be 256 $\times$ 256.
(iv) During training, multiplicative Gaussian noise was applied to unit weight values, with a standard deviation of 0.08, extracted from hardware measurements.
For the \ac{AIMC} chip and \ac{AIMC} chip model, the HWA trained weights were directly deployed.
To deploy weights on the digital hardware model (described below), \ac{PTQ} was performed on the HWA trained weights\footnote{PTQ was performed using the default options as described at \url{https://github.com/Xilinx/brevitas/tree/master/src/brevitas_examples/imagenet_classification/ptq} using \emph{flexml}.}. 

\subsection*{Hardware Models}
Two hardware models were used, representative of (i) a low-precision fixed-point digital hardware accelerator with 8-bit activations and 4-bit weights, and (ii) the IBM HERMES Project Chip~\cite{Y2023legalloNatElec} (a \ac{PCM}-based \ac{IMC} accelerator).

\subsubsection{Low-Precision Fixed-Point Digital Hardware Accelerator}
The operation of the low-precision fixed-point digital hardware accelerator was assumed to be deterministic. The Brevitas~\cite{brevitas} library was used to model inference of a low-precision fixed-point digital hardware accelerator.
Channel-wise quantization was performed where weight terms were quantized to a 4-bit fixed-point representation and activation terms were quantized to a 8-bit fixed-point representation.

\subsubsection{\ac{PCM}-based \ac{AIMC} Accelerator}
For some experiments, the IBM HERMES Project Chip\cite{Y2023legalloNatElec}, a 64-core \ac{AIMC} chip based on back-end integrated \ac{PCM} in a 14-nm \ac{CMOS} process with a total capacity of $4,194,304$ programmable weights, was used.
For others, when using this hardware was not possible, a representative hardware model was used instead.
The operation of this model was verified by matching experimental data from distinct noise sources and high-level behaviours from hardware experiments.
\ac{ADC} and \ac{DAC} circuits were assumed to operate ideally at 8-bit precision. Weight (programming) noise was modelled by fitting a third order polynomial function of the mean error (standard deviation) with respect to normalized weight values. For each normalized weight value, the error was assumed to be Gaussian distributed. Read noise was modelled using a lookup table, where the standard deviation (also assumed to be Gaussian distributed) of the read noise was recorded as a function of the conductance (in \ac{ADC} units). 
Finally, output noise was modelled by adding a Gaussian distributed additive noise, to each column for each mini-batch.
Other non-linear residual noise sources were assumed to be negligible.

\subsubsection{Modified \ac{PCM}-based \ac{AIMC} Accelerator}
We perform experiments to determine the dominant source of adversarial robustness. The aforementioned AIMC chip model was modified to consider only a single stochastic noise source.
To set the noise magnitude to produce a desired average test set accuracy value, Bayesian Optimization was performed using Optuna\cite{optuna}.
Specifically, 100 trails were executed with the Tree-structured Parzen Estimator (TPESampler) and the Median pruning algorithm (MedianPruner) optimization algorithms.

\subsection*{Adversarial Success Rate Evaluation}
As mentioned, to evaluate the adversarial success rate, we reply on the \ac{ASR} metric. In terms of accuracy, the \ac{ASR} can be determined using Algorithm \ref{alg:cap}.
\begin{algorithm}[H]
\caption{Determination of the Adversarial Success Rate (ASR) metric.}\label{alg:cap}
\begin{algorithmic}
\Require The model, \texttt{model}, clean inputs, $\mathbf{x}$, clean targets, $\mathbf{y}$, adversarial inputs, $\mathbf{x_{adv}}$
\Ensure The Adversarial Success Rate (ASR)
\State \texttt{total\_adversarial, correct\_adversarial = 0}
\For{\texttt{($x$,$y$,$x_{adv}$) in zip($\mathbf{x}$,$\mathbf{y}$,$\mathbf{x_{adv}}$)}} \Comment{Iterate over each batch of the clean inputs, clean targets, and adversarial inputs}
    \State \texttt{predicted\_clean = model(}$x$\texttt{), predicted\_adversarial = model(}$x_{adv}$\texttt{)} \Comment{Compute clean and adversarial predictions}
    \State \texttt{predicted\_adversarial = predicted\_adversarial[predicted\_clean == $y$]} \Comment{Mask adversarial predictions}
    \State \texttt{target\_adversarial = target[predicted\_clean == $y$]} \Comment{Mask adversarial targets}
    \State \texttt{total\_adversarial += len(target\_adversarial)}
    \State \texttt{correct\_adversarial += sum(predicted\_adversarial != target\_adversarial)} \Comment{Increment the number of masked, adversarial predictions that are \emph{not} matched with the desired, i.e., original, targets}
\EndFor
\State \Return \texttt{correct\_adversarial / total\_adversarial}
\end{algorithmic}
\end{algorithm}

\subsection*{Attack Generation and Evaluation}
To generate the PGD, Square, and OnePixel attacks, the \ac{ART} was used. To generate the \ac{GBDA} attacks, the following\footnote{\url{https://github.com/facebookresearch/text-adversarial-attack}.} code repository was used.
All validation samples were processed sequentially, i.e., to generate adversarial inputs, a unit batch size was used.
For each attack type, two attack parameters were varied -- directly relating to the attack magnitude and number of iterations.
To determine an appropriate range for each attack parameter, first, a large range for each parameter was explored.
The upper-bound was determined for each parameter by manually inspecting and labelling adversarial inputs.
Then, the parameter space for which the ground truth labels differed was avoided.
Finally, the lower-bound was determined for each parameter by manually inspecting contour plots of the adversarial success rate and determining where the rate of change of the adversarial success rate plateaued.

For all experiments, when adversarial inputs were presented, they were interleaved with their corresponding standard inputs. This was done for two reasons. First, to determine the \ac{ASR}, and second, to counteract temporal effects, such as drift.
All other attack parameters were set to default values\footnote{As of commit ID de99dca9e0482b43d2e5118f76d1b07135fcda51 (\url{https://github.com/Trusted-AI/adversarial-robustness-toolbox}).}.

For all target platforms, we consider the strongest scenario (i.e., the generation and evaluation of adversarial inputs using the same platform). Hence, all experiments consider four sets of network parameters\footnote{For all target platforms, it is not possible to represent the weights in the exact same representation, i.e., for the \ac{AIMC} chip and corresponding model, weights are encoded as conductance values, and for the original model they are represented using floating-point values.}: (i) the original floating-point weights, (ii) floating-point precision HWA retrained weights, (iii) HWA retrained, \ac{PTQ} weights, and (iv) HWA retrained weights mapped to conductance values.
For the AIMC chip, the mapped weights of the \ac{HWA} trained network were not used directly, as to mitigate mismatches in the assumed hardware configuration (during HWA retraining) and the \ac{AIMC} chip, a number of post-mapping calibration steps are performed.
We report the test set accuracy for all configurations (sets) of network parameters and platforms in Extended Data Table \ref{configuration_accuracy}.

\subsection*{Model Deployment and Inference for the \ac{AIMC} Chip}
Model deployment and inference for the \ac{AIMC} chip was performed using a sophisticated software stack, which allows for end-to-end deployment of \ac{DL} models.
First, using the {torch.fx}\cite{torch-fx} library, PyTorch models are traced using symbolic tracing, which propagates proxy objects through them. Graph-based representations of models are formed, where vertices represent operations and directed edges represent connections between sequences of operations.

Second, pipelining is performed at the operation level of abstraction. Operations are grouped into distinct chains and branches. Chains are comprised of one or more branches, which execute in parallel. Operations within branches execute sequentially. The inputs of the branches are either the output values of the previous chain or cached values stored previously during the execution.
At the first pipeline stage, new input is fed to the first chain and executed. At the second pipeline stage, new input is once again fed to the first chain and executed, however, the output(s) of the first chain is(are) fed to the second chain and executed. This process continues until the pipeline is full, and for each pipeline stage, all chains are executed. After the last input has been fed to the first chain, the pipeline is flushed until the final output has been received. This process maximizes the number of parallel \acp{MVM} performed on-chip.

Next, the floating-point weights of mapable layers (to \ac{AIMC} tiles) from the input \ac{DL} model are mapped to programmable conductance states, ranging from 0 - $G_{max}$, where $G_{max}$ is typically set to $160\mu S$.
Post-training optimizations~\cite{lammie2024improving} are then performed to tune the (i) input range of each \ac{AIMC} tile and the (ii) maximum conductance range of each column.
After these optimizations are performed, before the tuned target conductance states are programmed to the devices using \ac{GDP}~\cite{jub,10281389}, they are stored in floating-point precision to be deployed on other hardware platforms.
Finally, the software stack handles the execution of the compiled model by interfacing itself with lower-level drivers, which handle the execution of the operations on the \ac{AIMC} chip.

\subsubsection*{Hardware-in-the-loop Adversarial Attacks Targeting the AIMC Chip}
\color{black}
To perform hardware-in-the-loop attacks targeting the \ac{AIMC} chip, for the white-box \ac{PGD} attack, (i) the inputs and (ii) the weights of each layer, in addition to (iii) the outputs of the network, are required to compute gradients.
For inherently stochastic systems, this is problematic, as the programmed weights differ from the desired weights.
Moreover, weight values randomly fluctuate during inference operation due to high- and low-frequency weight noise, and conductance drift.
As described in Eq. (1), representative weights (associated with a given period of elapsed time) must be inferred by solving an optimization problem.
For the semi black-box OnePixel attack, probability labels and output logits are required as well.

For both of these types of attacks, before the attack was performed, \ac{HWA} weights were programmed to conductance values and post-training optimizations were performed, as described in \emph{Model Deployment and Inference for the AIMC chip}. 
For both of these types of attacks, adversarial inputs were generated by iteratively inferring inputs using the AIMC chip, performing backwards propagation, and using the resulting information to modify the adversarial inputs. Hence, the generation of one input in a sequence is dependent on the output of the previous input in a sequence. 
This is also problematic, as it means that these attacks cannot be pipelined. Consequently, they take a relatively long time to run, and are susceptible to conductance drift.
As a consequence, all hardware in the loop adversarial attacks targeting the AIMC chip are generated and evaluated for a randomly sampled subset (1,000/10,000) of the test set, where the number of samples in each class is equal (100). The same split was used for all associated experiments.

To perform both types of attacks, the following steps were performed.
First, initial inputs, as determined by the attack algorithm, and clean inputs, were constructed. These were propagated through the AIMC chip. During inference, for white-box attacks, the inputs to each layer were cached.
For the \ac{AIMC} chip, inputs and outputs of each layer can easily be cached using our software stack.
In a practical setting, voltage traces could be probed, or a side-channel attack could be performed to obtain these values.
Representative weight values were inferred by solving the optimization problem described in Eq. (1) by performing 2,000 \acp{MVM} on each crossbar, with inputs sampled from a clipped Gaussian distribution.
Again, in a practical setting, even if the attacker does not have direct control of the inputs and outputs of each crossbar, these could be measured during normal operation and Eq. (1) could be applied.
For both types of attacks, the output logits were also cached.

Next, the PyTorch~\cite{paszkePyTorchImperativeStyle2019} library was used to construct a proxy graph of the original network deployed on the \ac{AIMC} chip. Random inputs were propagated through this proxy graph to construct a computational graph used for back-propagation.
The cached inputs and weights to each layer were overridden using the cached values. The output logits were also overridden.
Gradients were computed using the auto-grad functionality of PyTorch on the proxy graph, and the gradients, in addition to the output logits, were fed to the attack algorithm to modify the current adversarial inputs to generate the next batch of inputs to infer.
This process was repeated for each attack iteration, and for each batch of inputs.
\color{black}

\newpage
\section*{Data availability}
The data that support the plots within this paper and other findings of this study are available from the corresponding authors upon reasonable request.

\section*{Code availability}
The code used to perform the simulations included in this study is available from the corresponding authors upon reasonable request.

\section*{Acknowledgments}
This work was supported by the IBM Research AI Hardware Center We would like to thank Benedikt Kersting, Irem Boybat, Hadjer Benmeziane, and Giacomo Camposampiero, for fruitful discussions. We would also like to thank Vijay Narayanan and Jeff Burns for managerial support.

\section*{Author contributions}
C. L. and A. S. initiated the project. C.L. designed and planned the project. C.L. and J.B. set up the infrastructure for generating and evaluating the adversarial attacks. C.L., J.B., G.C., and A.V. set up the infrastructure for automatically deploying trained models on the IBM Hermes Project Chip.
C.L., J.B., and A.V. wrote the manuscript with input from all authors. M.L.G. and A.S. supervised the project.

\section*{Competing interests}
The authors declare no competing interests. 
\clearpage

\renewcommand{\figurename}{FIG.}
\setcounter{figure}{0}
\clearpage
\begin{figure*}[!t]
\centering
\includegraphics[width=1\textwidth]{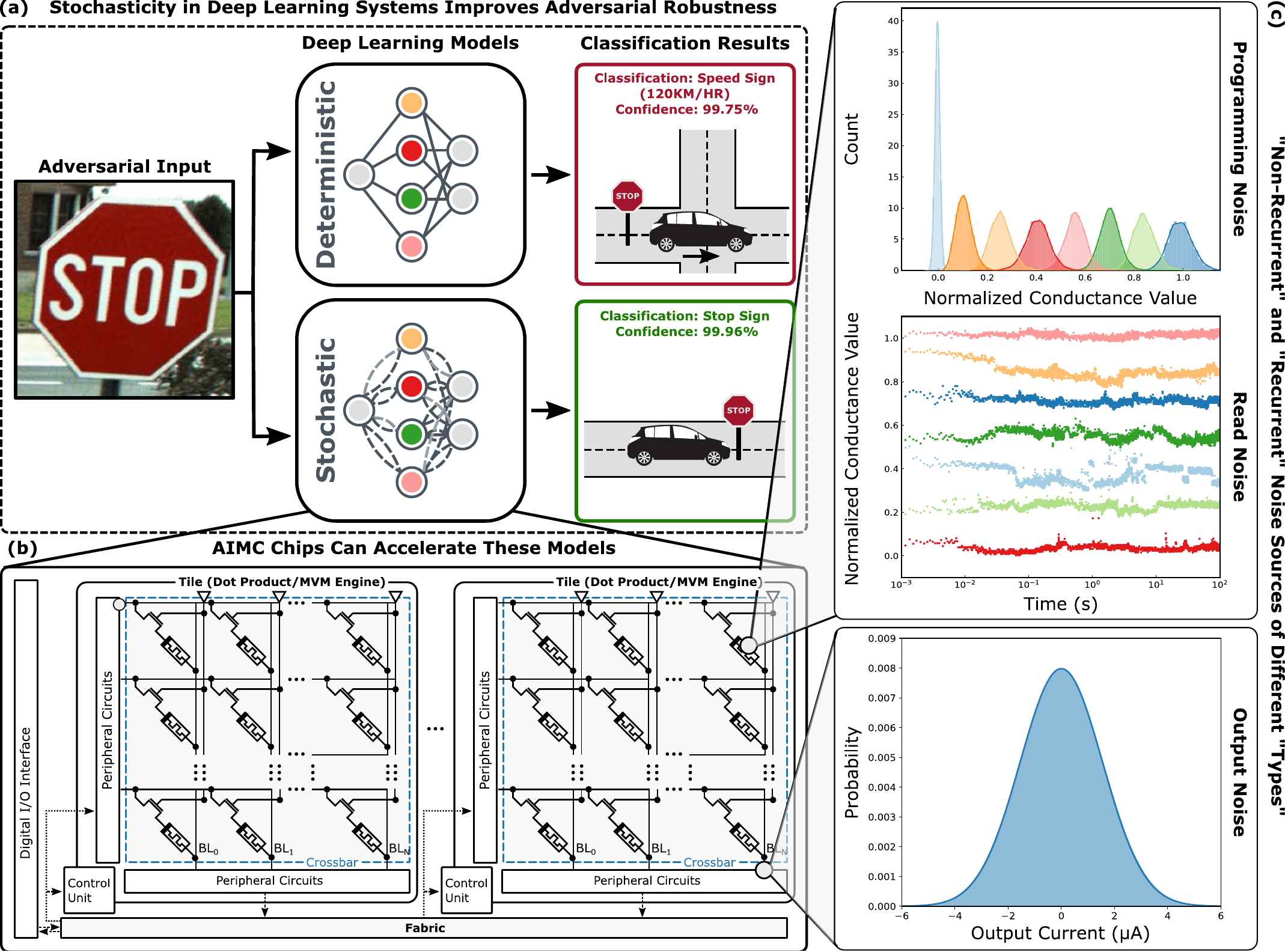}
\caption{
\textbf{Adversarial robustness of analog in-memory computing}. \textbf{(a)} An augmented (adversarial) image of a stop sign, intercepted and replaced by a malicious attacker, intended to be mis-categorized as a speed sign, is fed into a (i) deterministic and (ii) stochastic DNN accelerator residing in an autonomous vehicle. The deterministic system incorrectly identifies the input as a speed sign, whereas the stochastic system correctly identifies the input as a stop sign. \textbf{(b)} Analog in-memory computing chips which can be used to execute DNN inference workloads, are inherently stochastic. The circuits and devices they are constructed of introduce many different linear and non-linear \emph{noise sources}, including, but not limited to: quantization noise, weight programming noise, weight read noise, circuit noise, and temporal weight drift. \textbf{(c)} Noise sources have a number of distinct properties. The \emph{recurrence} property determines whether a noise source is "non-recurrent" or "recurrent". Non-recurrent noise sources introduce noise once (usually when the system is configured), while recurrent noise sources introduce noise at different frequencies during normal system operation.
The \emph{type} property determines whether the noise magnitude is determined as a function of the input. As depicted, programming noise is non-recurrent and its magnitude is input dependent. Read noise is recurrent and its magnitude is input dependent. We model output noise such that it is recurrent and its magnitude is input-independent.
}\label{fig:concept}
\end{figure*}

\clearpage
\begin{figure*}[!t]
\centering
\includegraphics[width=1\textwidth]{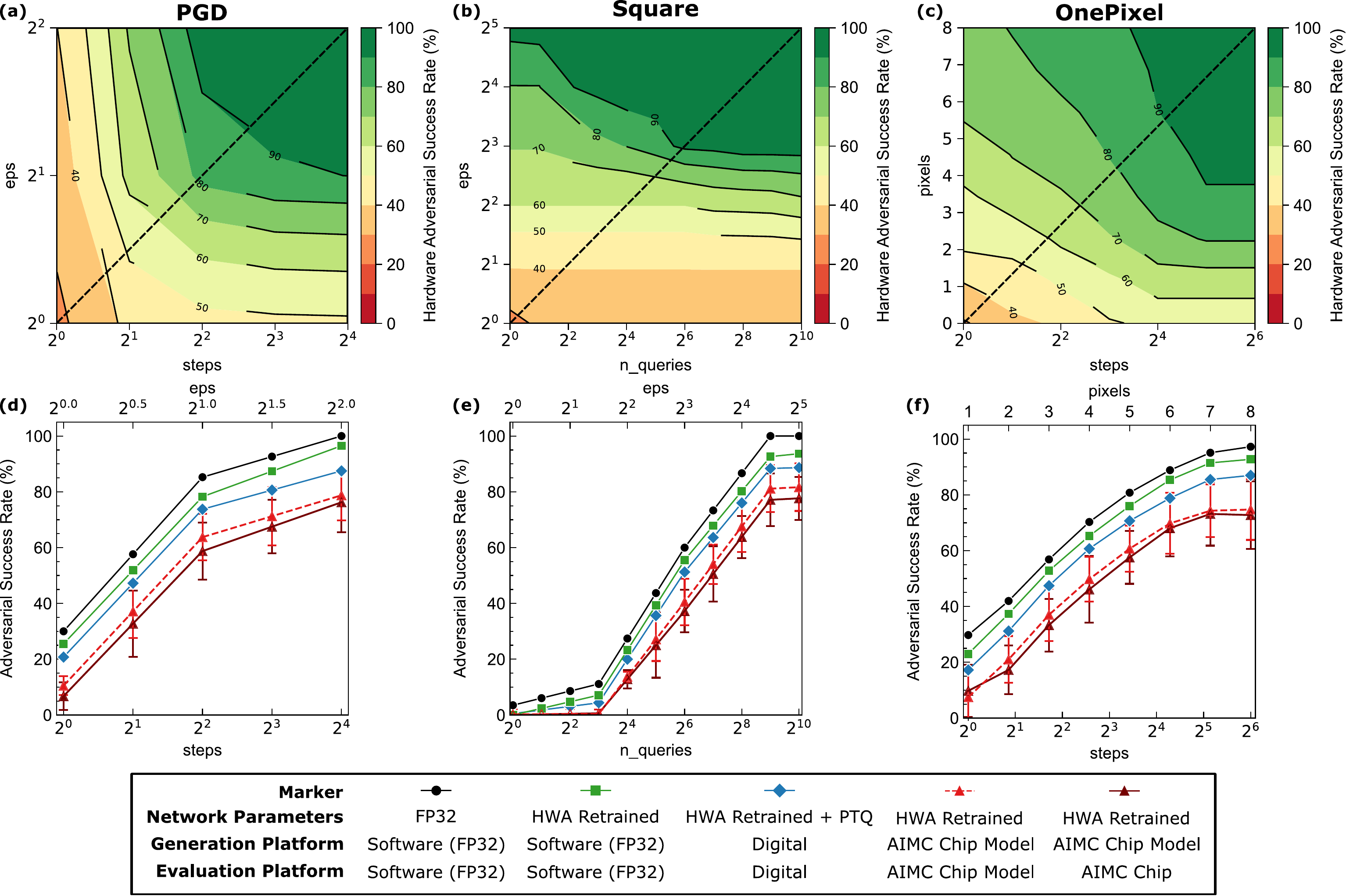}
\caption{
\textbf{Experimental validation of adversarial robustness}. \textbf{(a-c)} Adversarial inputs generated using the PGD, Square, and OnePixel adversarial attacks for different values of both attack parameters. The attacks are evaluated using the \ac{ASR} metric for the AIMC chip evaluation platform. \textbf{(d-f)} For different configurations, denoted using distinct marker and line styles, the ASR envelope (from the dashed lines in \textbf{(a-c)}) of each evaluation space is compared. For non-deterministic evaluation platforms, evaluation experiments are repeated $n=10$ times. Mean and standard deviation values are reported.
}\label{fig:hwexp}
\end{figure*}

\clearpage
\begin{figure*}[!t]
\centering
\includegraphics[width=1\textwidth]{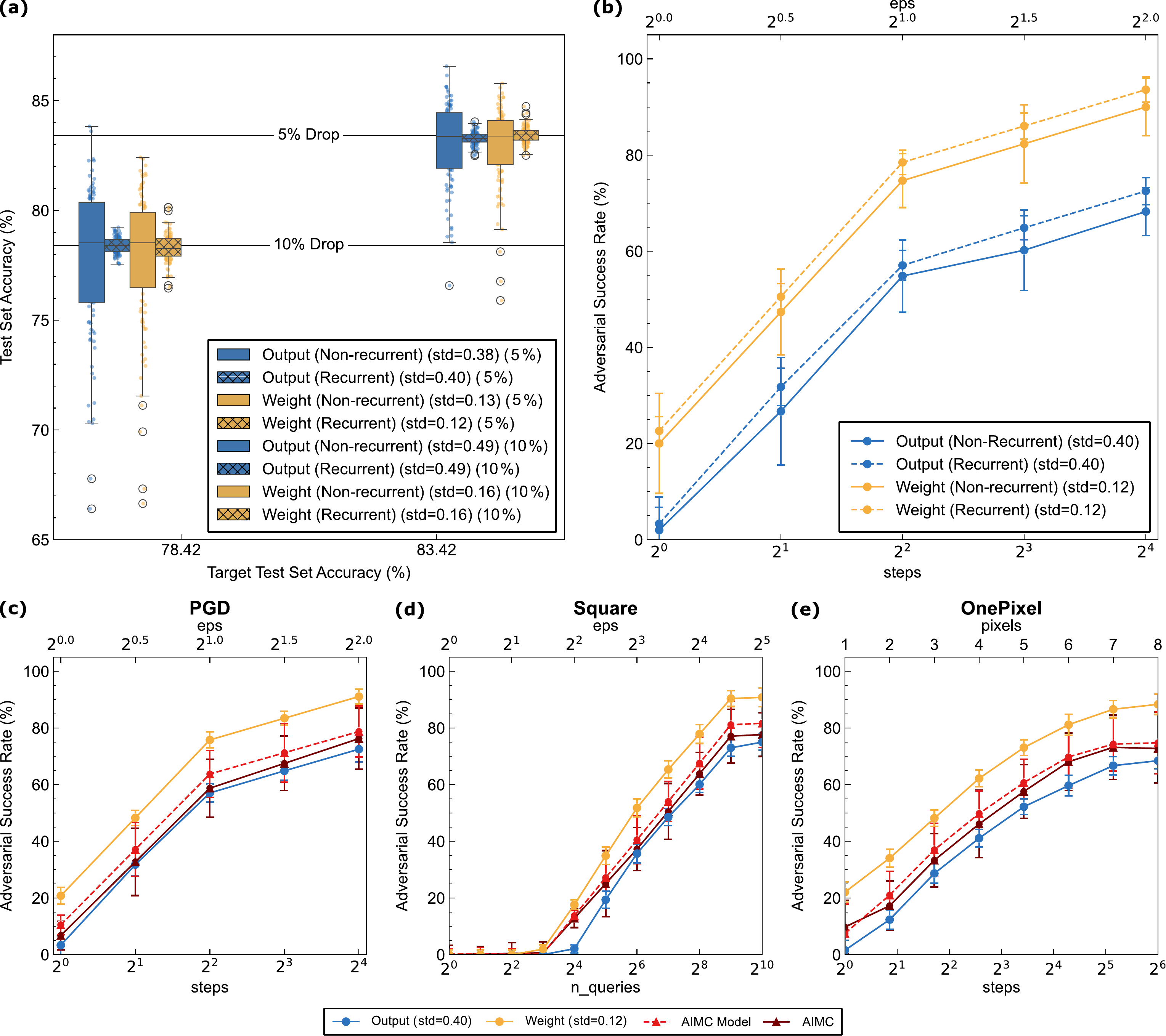}
\caption{
\textbf{The source of adversarial robustness.}
\textbf{(a)}
The test set accuracy for \ac{AIMC} models where only output or weight noise is considered, when the noise magnitude is set to result in test set accuracy drops (compared to the original floating-point model) of 5\% and 10\%. The test accuracy is evaluated for $n$=1,000 repetitions and mean and standard deviation values are reported \textbf{(b)} For PGD, the robustness of both non-recurrent and recurrent output and weight noise that result in a 5\% test set accuracy drop is evaluated. \textbf{(c-e)} The \ac{ASR} for the \ac{PGD}, Square, and OnePixel attacks are compared for the \ac{AIMC} models with only output and weight noise (both recurrent), the \ac{AIMC} model, and the AIMC chip. For non-deterministic evaluation platforms, evaluation experiments are repeated $n=10$ times. Mean and standard deviation values are reported.
}
\label{fig:source}
\end{figure*}

\clearpage
\begin{figure*}[!t]
\centering
\includegraphics[width=1\textwidth]{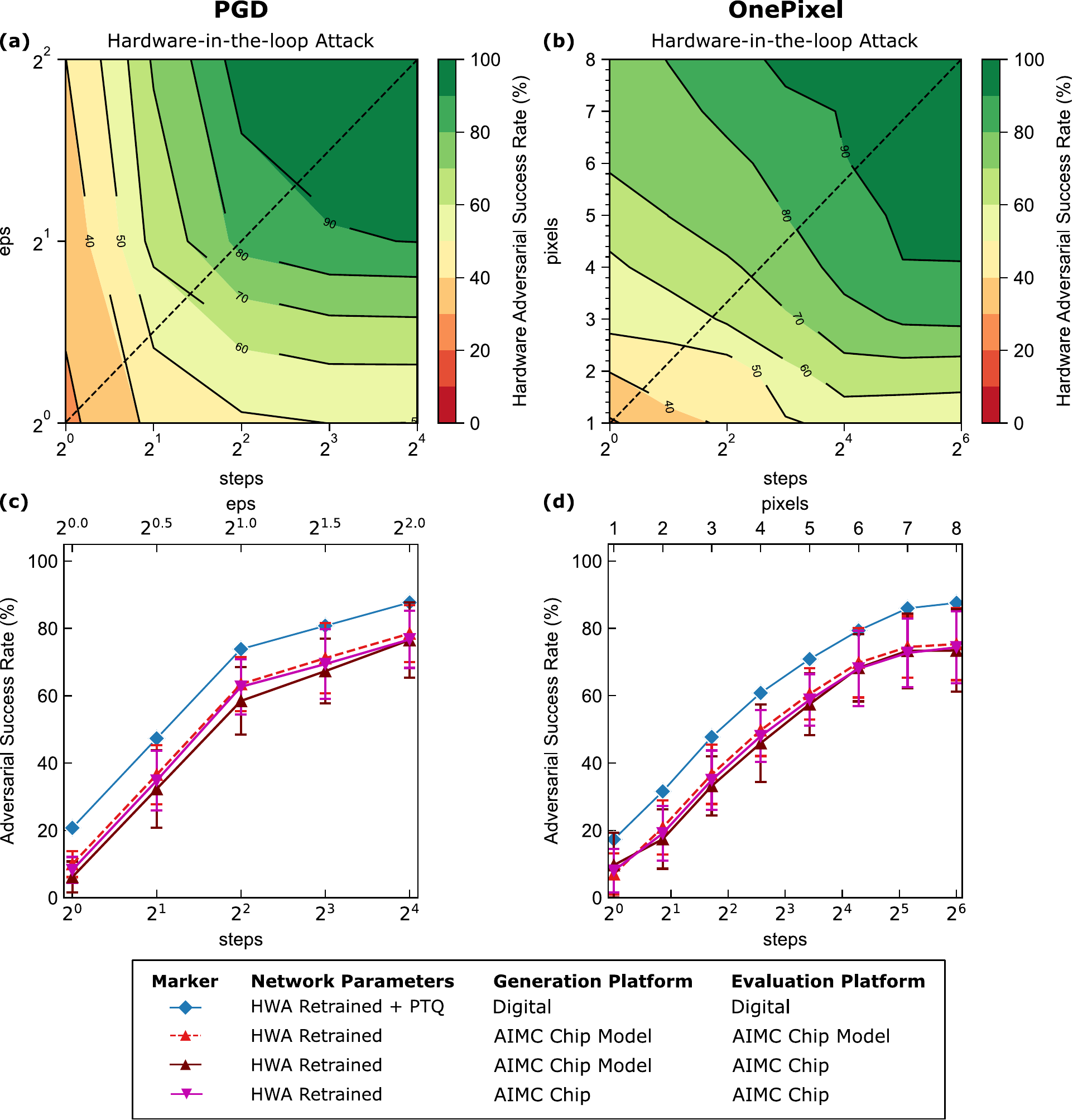}
\caption{
\textbf{Robustness to hardware-in-the-loop adversarial attacks.} 
\textbf{(a,b)} Adversarial inputs generated using the PGD and OnePixel adversarial attacks, both generated and evaluated using the AIMC chip.  \textbf{(c,d)} When evaluated using the AIMC chip, adversarial inputs generated using hardware-in-the-loop attacks are more effective than when they are generated using the AIMC chip, compared to when they are generated using the AIMC chip model. Adversarial inputs generated and evaluated using the AIMC chip model exhibit similar adversarial robustness to the AIMC chip. For all three configurations, additional adversarial robustness is observed compared to when they are generated and evaluated using the digital hardware accelerator. For non-deterministic evaluation platforms, evaluation experiments are repeated $n=10$ times. Mean and standard deviation values are reported.
}\label{fig:hwloop}
\end{figure*}

\clearpage
\begin{figure*}[!t]
\centering
\includegraphics[width=1\textwidth]{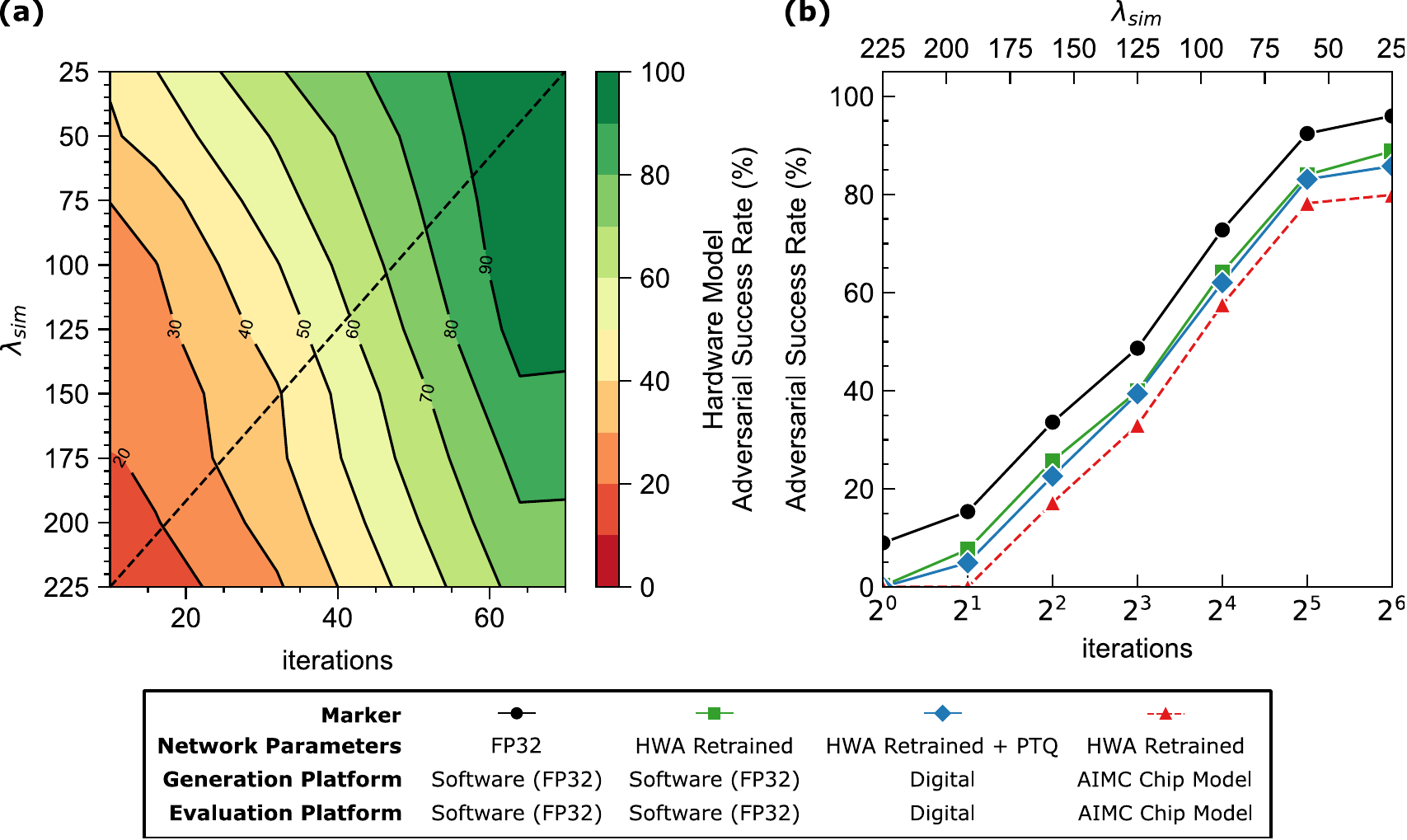}
\caption{
\textbf{Robustness of transformer-based models and natural language processing tasks to adversarial attacks.} 
\textbf{(a)} Using the \ac{GBDA} adversarial attack, adversarial inputs are both generated and evaluated using the AIMC chip model.  The \ac{ASR} is reported for different values of both attack parameters (iterations and $\lambda_{sim}$).
\textbf{(b)} The ASR envelope is reported for a number of strongest attack configurations, which are denoted using distinct marker and line styles, the ASR envelope (from the dashed lines in \textbf{(a-c)}) of each evaluation space is compared.
}\label{fig:transformer}
\end{figure*}

\clearpage
\renewcommand{\tablename}{SUPPLEMENTARY TABLE.}
\newpage
\section*{Supplementary Tables}
\begin{table}[H]
\centering
\caption{The test set accuracy of different configurations (platform and network parameters).}\label{configuration_accuracy}
\begin{tabular}{llr} \toprule \toprule
\textbf{Platform} & \textbf{Network Parameters} & \textbf{Test Set Accuracy (\%)} \\ \midrule
\multicolumn{3}{c}{\textbf{Resnet-based CNN (CIFAR-10)}} \\ \midrule
Software (FP32) & FP32 & 88.42 \\
Software (FP32) & HWA Retrained & 87.80 \\
Digital & HWA
  Retrained + PTQ & 86.56 \\
Analog
  Model & HWA Retrained & 84.85 \\
Analog Hardware & HWA Retrained & 84.31 \\ \midrule
\multicolumn{3}{c}{\textbf{RoBERTa (MNLI)}} \\ \midrule
Software 
  (FP32) & FP32 & 86.20 \\
Software (FP32) & HWA Retrained & 85.72 \\
Digital & HWA
  Retrained + PTQ & 82.69 \\
Analog
  Model & HWA Retrained & 81.96 \\ \bottomrule \bottomrule
\end{tabular}
\end{table}

\begin{table}[H]
\centering
\caption{Dependence of the \acl{ASR} metric on the test set accuracy for a floating-point Resnet-based CNN trained for CIFAR-10 image classification using two different attacks.}\label{independence_table}
\begin{tabular}{lccc}
\toprule \toprule
\multirow{2}{*}{\textbf{Epoch}} & \multirow{2}{*}{\textbf{Test Set Accuracy (\%)}} & \multicolumn{2}{c}{\textbf{ASR (\%)}} \\
 &  & \textbf{PGD (steps=2, eps=2)} & \textbf{OnePixel (stpes=4, pixels=3)} \\ \midrule
1 & 21.98 & 77.45 & 71.17 \\
10 & 64.46 & 86.31 & 64.03 \\
20 & 72.88 & 85.59 & 64.65 \\
30 & 77.18 & 86.08 & 65.24 \\
40 & 79.02 & 85.27 & 64.76 \\
50 & 80.84 & 85.38 & 64.78 \\
100 & 88.39 & 85.34 & 64.81 \\
\bottomrule \bottomrule
\end{tabular}
\end{table}

\clearpage
\begin{table}[H]
\centering
\caption{Original and adversarial text inputs for different $\lambda_{sim}$ and $n_{iters}$ parameters.}\label{table:prompts}
\begin{threeparttable}
\begin{tblr}{
  width = \linewidth,
  colspec = {Q[123,l]Q[173,l]Q[175,l]Q[90,c]Q[117,c]Q[90,c]Q[163,c]},
  cell{1}{1} = {c=2}{c},
  cell{1}{3} = {r=2}{},
  cell{1}{4} = {r=2}{},
  cell{1}{5} = {r=2}{},
  cell{1}{6} = {r=2}{},
  cell{1}{7} = {r=2}{},
  cell{2}{1} = c,
  cell{2}{2} = c,
}
\toprule \toprule
\textbf{Sentence 1} &  & \textbf{Sentence 2} & \textbf{Label} & \textbf{$\lambda_{sim}$} & \textbf{$N_{iters}$} & \textbf{BERTScore} \\
\textbf{Original} & \textbf{Adversarial} &  &  &  &  & \\ \midrule
Ah, ma foi, no! replied Poirot frankly. & Ah, ma foi, mage! replied Poirot frankly. & Poirot agreed with what I just said. & contradiction & 25 & 1 & 0.96\\ \midrule

Do you want to see historic sights and tour museums and art galleries? & Dodo want go see see sights and tour traditional,, movie?. & Would you like to visit historic places, museums, and art galleries? & entailment & 50 & 16 & 0.89 \\ \midrule

I regretfully acknowledge that it may even make practical sense to have a few hired guns like Norquist, Downey, and Weber around--people of value only for their connections to power, not for any knowledge or talent. & I regretfully acknowledge that it may even make practical sense to have a agreement hired guns like Norquistait Downey, and hi Katy--peoplecond value only for their connections life power, not for any knowledge or talent. & It might be a good idea to have hired guns around. & entailment & 75 & 32 & 0.95 \\ \midrule

Outside the cathedral you will find a statue of John Knox with Bible in hand. & OutsideBeyond streets landmark will find found some monument John Sir with biblical bible hand handle. & There are many statues in front of the cathedral of famous religious people. & neutral & 100 & 64 & 0.91 \\ \midrule

Something may be better than nothing. If trials compared low-cost therapy to the complete AZT regimen it's likely that the new regimens will prove less effective. & Something perhaps be better than nothing. If technology compared low-cost therapy to the complete AZ15 regimen it it likely that that new regimensimens prove less successful successful. & It is better to have little than to have none at all. & entailment & 200 & 128 & 0.94\\
\bottomrule \bottomrule
\end{tblr}
\begin{tablenotes}
\item[1] Average for both sentences.
\end{tablenotes}
\end{threeparttable}
\end{table}

\newpage
\renewcommand{\figurename}{SUPPLEMENTARY NOTE FIG.}
\renewcommand{\tablename}{SUPPLEMENTARY NOTE TABLE.}
\color{black}
\section*{Supplementary Note 1: AIMC chip model}
\setcounter{figure}{0}

\begin{figure*}[!h]
    \includegraphics[width=1\textwidth]{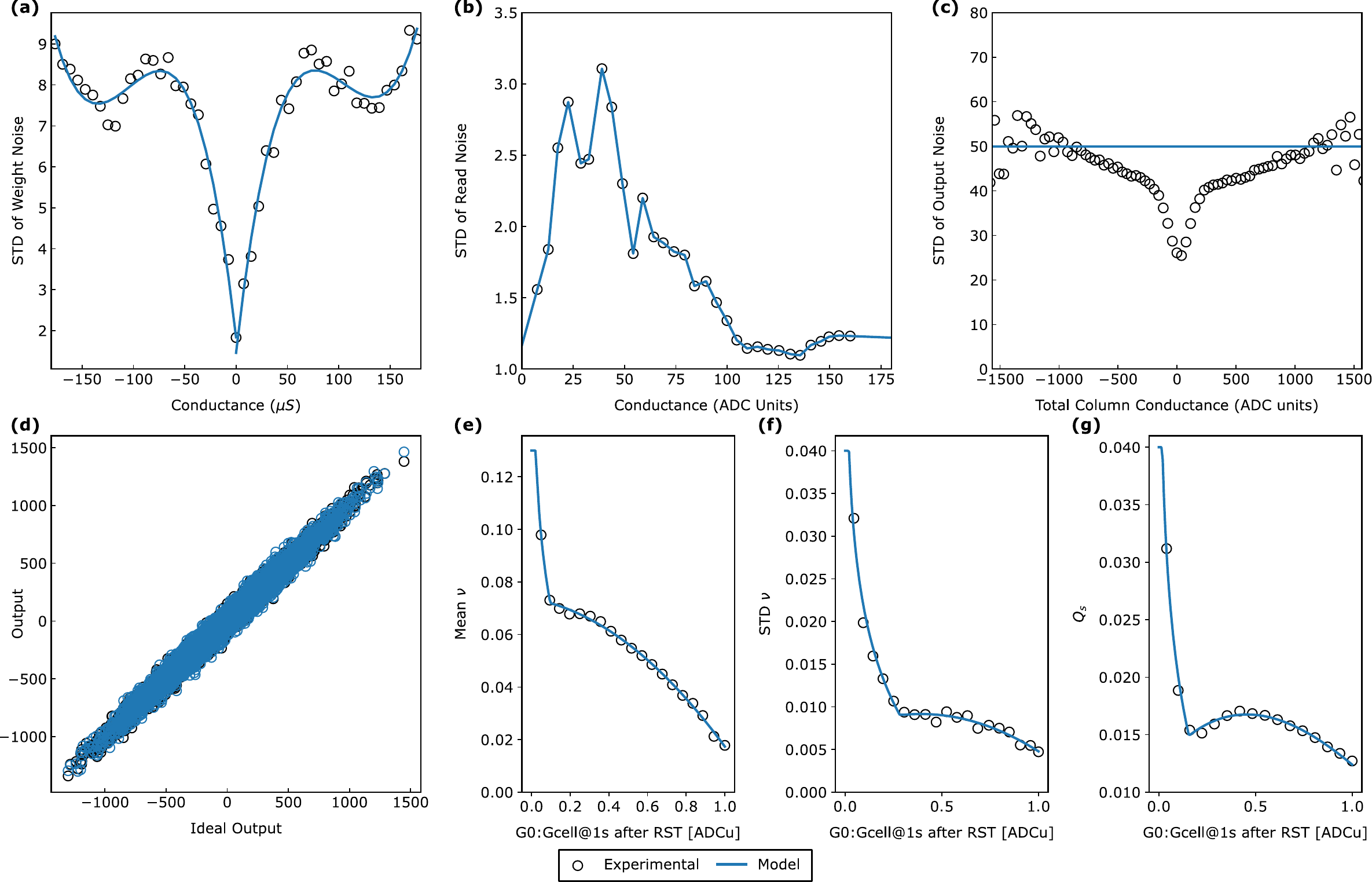}
    \caption{\color{black}\textbf{AIMC chip model}. Match for the (a) weight noise, (b) read noise, (c) output noise, and (d) matrix-vector multiply operation. For weight noise, a third-order polynomial fit is used. For read noise, experimental data is interpolated and a look-up table is used. For output noise, a fixed standard deviation is used (assumed to be independent of the total column conductance). One ADC unit corresponds to approximately 0.115 $\mu$S. (e) Mean of the drift coefficient as a function of normalized (to Gmax) ADC unit value. (d) Standard deviation of the drift coefficient as a function of the normalized (to Gmax) ADC unit value.}
    \label{hwmatch}
\end{figure*}

The construction and verification of the PCM-based \ac{AIMC} chip model, which was used for some simulations, was performed using experimental data from hardware experiments. In Supplementary Note Fig. \ref{hwmatch}, the match between experimental data and the model is depicted for weight, read, and output noise, in addition to matrix-vector multiply operations and temporal drift -- for two-device pairs -- which are used to encode weight components in the IBM HERMES project chip.

To apply drift noise to the \ac{AIMC} chip model, read noise is scaled by the ratio of the current device conductance to the original, programmed device conductance. We refer the reader to Section IV of the Supplementary material of~\cite{Gallo2022bs} for the exact methodology employed to measure the mean and standard deviation of the drift coefficient.

\newpage
\section*{Supplementary Note 2: Adversarial Robustness to Varying Degrees of Stochasticity}

\begin{table}[!h]
\color{black}
\centering
\caption{\color{black}Severity (standard deviation) of two isolated stochastic noise sources: output and weight noise, which yield
test set accuracy drop values$^1$ between 1\% and 20\% for the Resnet-based \ac{CNN}.}\label{table:severity}
\begin{threeparttable}
\begin{tabular}{lccccc}
\toprule \toprule
\begin{tabular}[c]{@{}l@{}}\textbf{Noise Type/Test Set Accuracy}\\\textbf{Drop Threshold (\%)}\end{tabular} & 1 & 2.5 & 5 & 10 & 20 \\
\midrule
Output
  (Non-recurrent) & 0.279 & 0.351 & 0.389 & 0.47 & 0.631 \\
Output
  (Recurrent) & 0.285 & 0.349 & 0.41 & 0.493 & 0.665 \\
Weight
  (Non-current) & 0.084 & 0.087 & 0.123 & 0.166 & 0.252 \\
Weight
  (Recurrent) & 0.076 & 0.082 & 0.125 & 0.164 & 0.24 \\
  \bottomrule \bottomrule
\end{tabular}
\begin{tablenotes}
\item[1] With respect to the floating-point accuracy, when clipped floating-point weights were used.
\end{tablenotes}
\end{threeparttable}
\end{table}

\begin{figure}[!h]
    \centering
\includegraphics[width=1\linewidth]{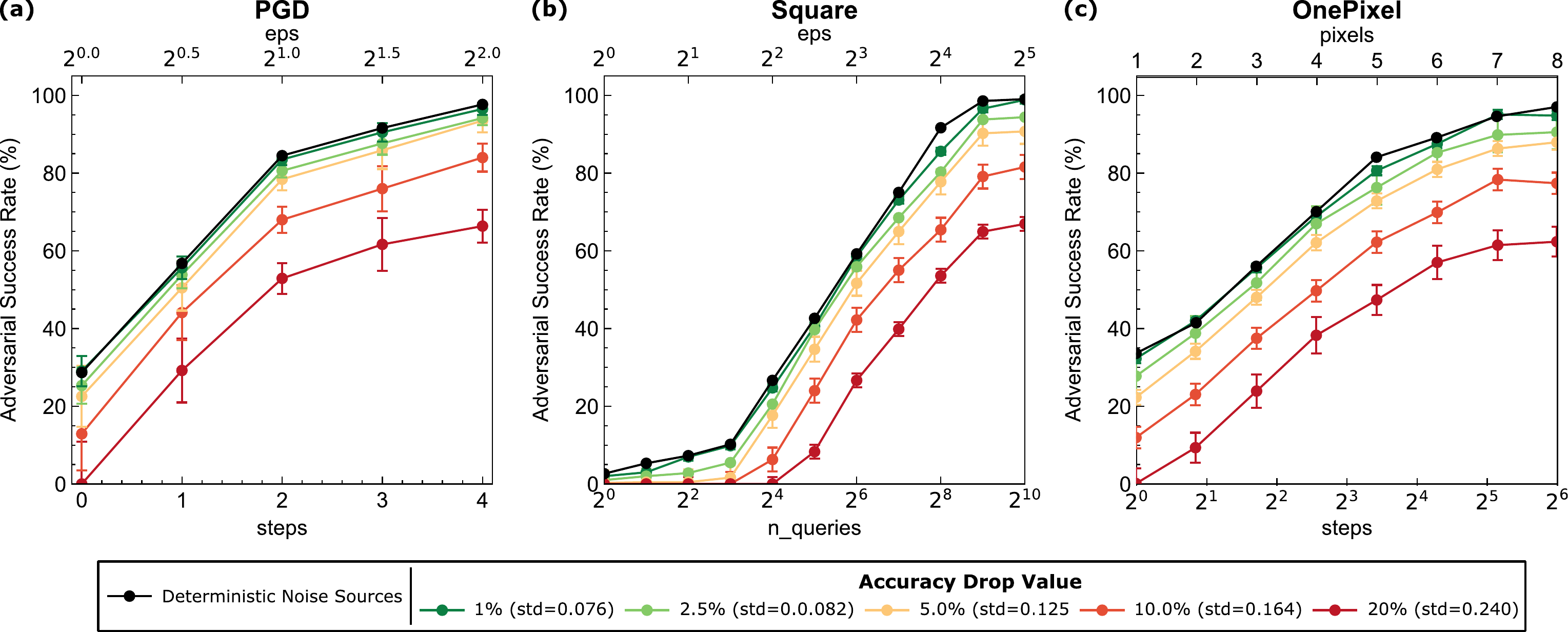}
    \caption{\color{black}\textbf{Adversarial robustness of the Resnet-based \ac{CNN} to varying degrees of stochasticity. (a-c)} The \ac{ASR} for the PGD, Square, and OnePixel attacks for the \ac{AIMC} model, where output noise is disabled and recurrent weight noise is modelled, resulting in test set accuracy drop values between 1 and 20\%. The adversarial robustness of the deterministic noise sources is also evaluated. Evaluation experiments are repeated $n=10$ times. Mean and standard deviation values are reported.}
    \label{fig:additional_accuracy_analysis}
\end{figure}

To further investigate how the degree of stochasticity effects the adversarial robustness of \ac{AIMC}-based hardware, in Supplementary Note Table~\ref{table:severity}, we determine the severity of two isolated stochastic noise sources: output and weight noise, which yield test set accuracy
drop values of 1\%, 2.5\%, 5\%, 10\%, and 20\%.
As a proxy for recurrent- and non-recurrent output noise, and non-recurrent weight noise, we simulate recurrent-weight noise.
In Supplementary Note Fig.~\ref{fig:additional_accuracy_analysis}, for the Resnet-based \ac{CNN}, we determine the \ac{ASR} envelope for recurrent-weight noise using the \ac{PGD}, Square, and OnePixel attacks. Additionally, we quantify and compare the adversarial robustness of the deterministic noise sources.

Intuitively, from Supplementary Note Fig.~\ref{fig:additional_accuracy_analysis}, it can be observed that as the test set accuracy drop threshold decreases, the \ac{ASR} increases. Conversely, as the test set accuracy drop threshold increases, the \ac{ASR} decreases. When only recurrent weight noise is modelled, a standard deviation of $\geq$ 0.125 is required to achieve the same degree of adversarial robustness as the \ac{AIMC} chip model and \ac{AIMC} chip. Deterministic noise sources contribute negligibly, exhibiting a small amount of additional robustness compared to the Software (FP32) model.

Performing this analysis enables us to observe of the trade-off between adversarial robustness and performance (accuracy).
As can be seen, in Supplementary Note Fig.~\ref{fig:additional_accuracy_analysis}, the \emph{average} adversarial robustness is not linearly proportional to the test set accuracy drop value. Small amounts of noise still result in some additional adversarial robustness.
In fact, when only one stochastic noise source is modelled, for drop values of 1\% and 2.5\%, which are more indicative of a realistic scenario, additional adversarial robustness is still observed -- just to a much smaller degree than that exhibited by the \ac{AIMC} chip and when output noise is modelled.
This same trend, i.e., the non-linear relation between the drop threshold and adversarial robustness, is expected to be observed for output noise.

\newpage
\section*{Supplementary Note 3: Simulations Using a More Challenging Image Classification Task}

\begin{table}[!h]
\centering
\color{black}
\caption{\color{black}The top-1 test set accuracy of different configurations (platform and network parameters) for the WRN-50-2-bottleneck network on the iNaturalist 2021 dataset.}\label{table:inaturalist_accuracy}
\begin{tabular}{lll}
\toprule \toprule
\textbf{Platform} & \textbf{Network Parameters} & \textbf{Test Set Accuracy (\%)} \\ \midrule
Software (FP32) & FP32 & 60.09 \\
Software
  (FP32) & HWA Retrained & 59.31 \\
Digital & HWA Retrained + PTQ & 58.27 \\
Analog
  Model & HWA Retrained & 57.66 \\ \bottomrule \bottomrule
\end{tabular}
\end{table}

\begin{figure}[!h]
    \centering
\includegraphics[width=1\linewidth]{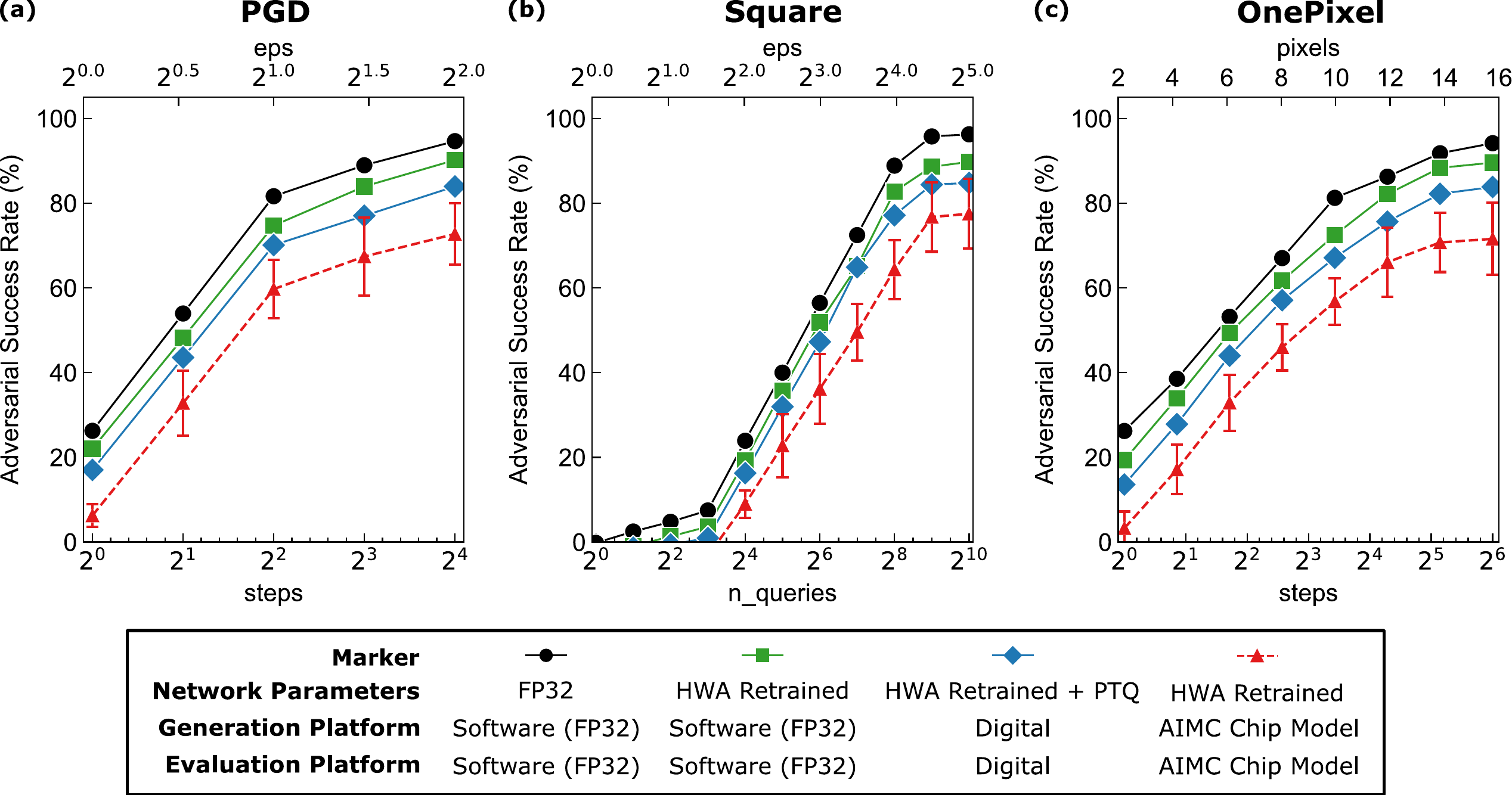}
\caption{\color{black}\textbf{Adversarial robustness of the WRN-50-2 bottleneck \ac{CNN} for the iNaturalist 2021 dataset. (a-c)} For different configurations, denoted using distinct marker and line styles, the ASR envelope of each evaluation space is compared. For non-deterministic evaluation platforms, evaluation experiments are repeated n = 10 times. Mean and standard deviation values are reported.}
\label{fig:inaturalist}
\end{figure}

To investigate the adversarial robustness for a larger image classification task, we adopt the iNaturalist~\cite{Horn2021} 2021 dataset and use the WRN-50-2-bottleneck~\cite{Zagoruyko2016} network.
The iNaturalist 2021 dataset contains a total of 10,000 plant and animal species represented by images of various sizes.
\ac{HWA} training was performed as described in the Methods section for the ResNet9S network with the following modifications. A batch size of 256 was used with the Adam~\cite{Kingma2015} optimizer. A step learning rate schedule was used with a step size of 50 epochs and $\gamma$=0.1.
A total of 500 training epochs were performed.
We report the top-1 test set accuracy of different configurations in Supplementary Note Table~\ref{table:inaturalist_accuracy}. The top-1 test set accuracy of the Software (FP32) model is inline with prior work which uses CNN-based networks with a similar number of parameters~\cite{Horn2021}.

As can be observed, in Supplementary Note Fig.~\ref{fig:inaturalist}, similarly to as in Fig. 2 in the main text, the \ac{AIMC} chip model has the smallest \ac{ASR} envelope, meaning that it is the most adversarially robust, for all three attack types (PGD, Square, and OnePixel).
A similar relationship between the adversarial robustness of the different platforms is also observed.
This is in an interesting result, as it indicates that for larger and more complex image classification tasks, and hence, more complex networks, additional adversarial robustness is still observed for the \ac{AIMC} chip model.

\newpage
\section*{Supplementary Note 4: Robustness to Temporal Drift}

\begin{figure}[!h]
    \centering
\includegraphics[width=1\linewidth]{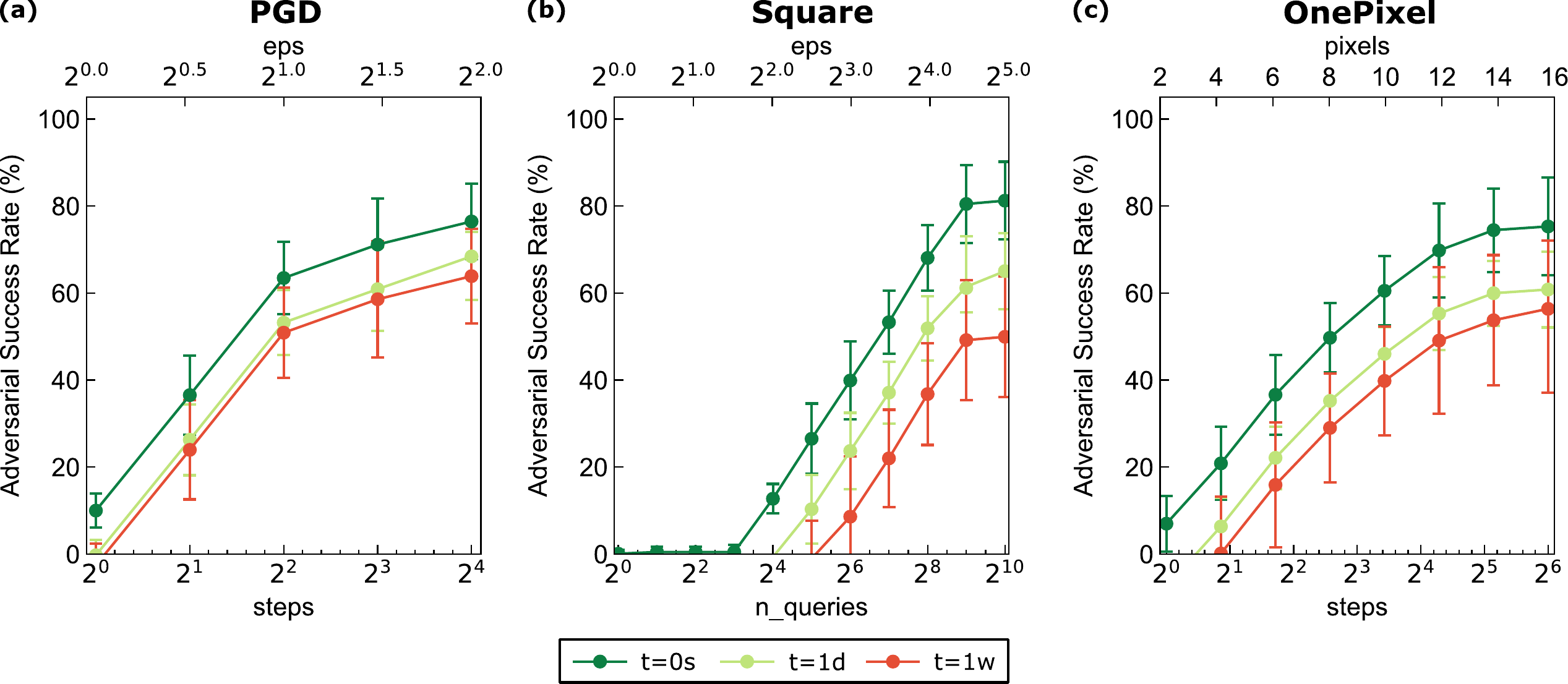}
\caption{\color{black}\textbf{Adversarial robustness of the Resnet-based CNN deployed on the \ac{AIMC} chip model with HWA trained weights as function of temporal drift. (a-c)} For different times, denoted using distinct line colors, the ASR envelope of each evaluation space is compared. Evaluation experiments are repeated n = 10 times. Mean and standard deviation values are reported.}
\label{fig:drift}
\end{figure}

\begin{table}[!h]
\color{black}
\centering
\caption{\color{black}The test set accuracy of the Resnet-based CNN on the CIFAR-10 dataset as a function of temporal drift. Evaluation experiments are repeated n = 10 times. Mean and standard deviation values are reported.}\label{table:drift_acc}
\begin{tabular}{lr}
\toprule \toprule
\textbf{Time (s)} & \textbf{Test Set Accuracy (\%)} \\
\midrule
0 & 87.06 $\pm$ 0.28 \\
86400 & 83.71 $\pm$ 0.52 \\
604800 & 80.87 $\pm$ 0.62 \\
\bottomrule \bottomrule
\end{tabular}
\end{table}

To investigate how the inherent adversarial robustness of \ac{AIMC}-based hardware is effected by temporal drift, and evolves over time, in Supplementary Note Fig.~\ref{fig:drift}, we perform additional simulations using the Resnet-based CNN, based off the drift data described in the supplementary note entitled \textit{AIMC Chip Model}.
Additionally, in Supplementary Note Table~\ref{table:drift_acc}, we report the mean and standard deviation of the resulting test set accuracy for 10 iterations.

It is well known~\cite{suriImpactPCMResistancedrift2013} that the compounding effects of temporal drift increase linearly with respect to logarithmic time. As can be observed in Supplementary Note Fig.~\ref{fig:drift}, while some variation between attack types is observed, the largest \emph{delta} in adversarial robustness is observed between t=0s and t=1d.
Clearly, effective robustness is still present over time, and even tends to increases as a function of drift time.

\newpage
\section*{Supplementary Note 5: Combinations of Output and Weight Noise}

\begin{figure}[!h]
    \centering
\includegraphics[width=1\linewidth]{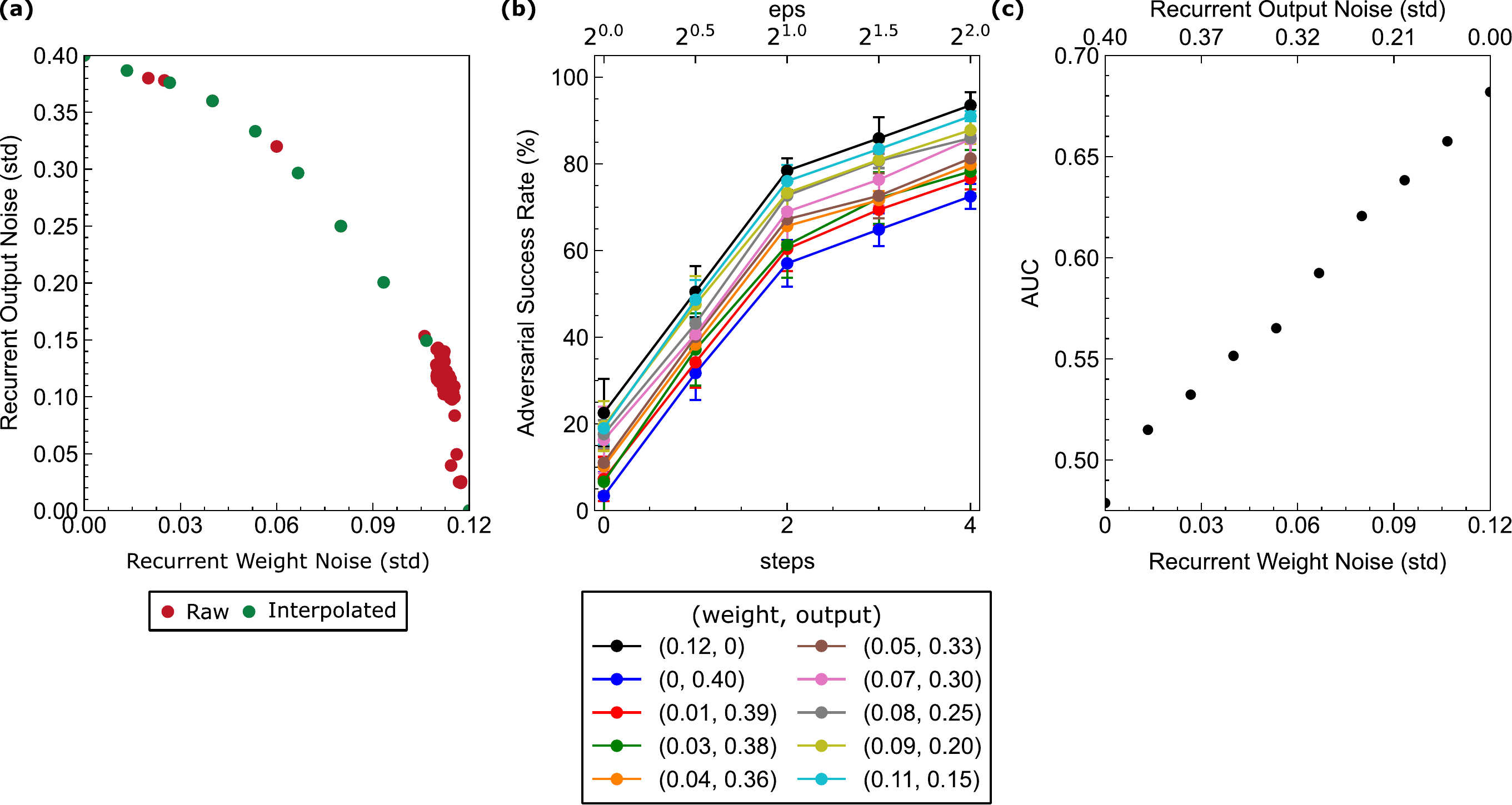}
\caption{\color{black}\textbf{Interactions between output and weight noise on adversarial robustness} for the Resnet-based CNN and PGD attack. (a) Combinations of recurrent output and weight noise which result in an accuracy drop of approximately 5\%. (b) For each (interpolated) combination, the resulting \ac{ASR} evelope. (c) For linearly increasing recurrent weight noise severity values, the area under each of the the \ac{ASR} envelopes.}
\label{fig:comb}
\end{figure}

To explore the interactions between different types of noise and their combined effects on robustness, we perform additional simulations using a modified version of the \ac{AIMC} chip model with the Resnet-based CNN and the \ac{PGD} attack. 
As was previously demonstrated in the original manuscript, recurrent and non-recurrent output and weight noise sources result in approximately the same adversarial robustness. Consequently, we perform simulations using different combinations of recurrent output and weight noise.

First, we determined to what degree the output and weight noise severity effects the test set accuracy. To do this, we varied the severity of both output and weight noise, and for a large number of potential combinations, evaluated the test set accuracy on the CIFAR-10 dataset.
All combinations which did not result in a test set accuracy drop of approximately 5\%, i.e., between 4.9\% and 5.1\%, were discarded, so that additional simulations could be performed for only combinations of output and weight noise severity values which yielded the same fixed test set accuracy drop value.

The non-discarded values (combinations of output and weight noise) are reported using red markers in Fig.~\ref{fig:comb}a.
Due to the relative scarcity of these points in particular regions, namely when the proportion of weight noise is relatively small, we performed linear interpolation to find other combinations of output and weight noise that have a linearly spaced proportion of weight noise.
These values also yielded test set accuracy drop values within the aforementioned acceptance bound. These values are reported using green markers in Fig.~\ref{fig:comb}a.

For each of these values, we determined their adversarial robustness (Fig.~\ref{fig:comb}b). As many of the \ac{ASR} envelopes are close in proximity, it is difficult to interpret their adversarial robustness relative to each other. Consequently, to aid comparisons, in Fig.~\ref{fig:comb}c, we compare the \emph{area} under each of the \ac{ASR} envelopes\footnote{\color{black}{Relative to the maximum area, calculated by multiplying the maximum number of steps by the maximum adversarial success rate (100\%).}}. A roughly linear relationship between the \ac{ASR} and the proportion of weight noise is observed, meaning that the proportion of output noise is inversely exponentially related to the \ac{ASR}. We conclude, that, even in the presence of both noise sources, the proportion of output noise has a much larger influence on adversarial robustness compared to weight noise.

\end{document}